\documentclass[amsmath, amssymb, preprintnumbers, showpacs, showkeys, aps, prl,superscriptaddress,twocolumn]{revtex4-2}
\usepackage{amssymb}
\usepackage[utf8]{inputenc}
\usepackage{graphicx}
\usepackage{amsmath}
\usepackage{csquotes}
\usepackage{xcolor}
\usepackage{placeins}
\usepackage{braket}
\usepackage{comment}
\usepackage[colorlinks, linkcolor={red!80!black},citecolor={blue!70!black},urlcolor={blue!80!black}]{hyperref}

\newcommand{\SM}{Supplemental Material}

\begin{document}

\title{Hopper-Like Growth of Higher-Order Topological Insulators}

\author{Yutaro Tanaka}
\email{yutaro.tanaka.ay@riken.jp}
\affiliation{
		RIKEN Center for Emergent Matter Science, Wako  351-0198, Japan
	}
    
\author{Shuai Zhang}
\affiliation{
	Institute of Theoretical Physics, Chinese Academy of Sciences, Beijing 100190, China
}

\author{Tiantian Zhang}
\affiliation{
	Institute of Theoretical Physics, Chinese Academy of Sciences, Beijing 100190, China
}

\author{Shuichi Murakami}
\email{murakami@ap.t.u-tokyo.ac.jp}
\affiliation{Department of Applied Physics, The University of Tokyo, Tokyo 113-8656, Japan}
\affiliation{
RIKEN Center for Emergent Matter Science, Wako 351-0198, Japan
	}
\affiliation{International Institute for Sustainability with Knotted Chiral Meta Matter (WPI-SKCM${}^\text{2}$),
Hiroshima University, Hiroshima 739-8526, Japan}

\begin{abstract}
Understanding crystal growth and morphology is a fundamental issue in condensed matter physics. 
While crystal morphology due to the distribution and dynamics of the diffusion field has been intensively studied, how the intrinsic material properties affect crystal morphology remains unclear.
In this Letter, we demonstrate that higher-order topological phases can give rise to hollowed crystal morphologies, where the corners advance faster than the central regions of the crystal, through an unconventional mechanism originating from topological electronic states. 
We quantitatively show this connection by analyzing both the fractal dimension $D_f$ and the fractal dimension of coastlines $D_{f,c}$. 
When we compare the crystals in the normal insulator and higher-order topological insulator phases with the same $D_{f}$ in the case of relatively rapid crystal growth, the former is in the dendritic shape, while the latter is in the hopper-like shape, quantified by the smaller $D_{f,c}$ in the higher-order topological phase. 
\end{abstract}

\maketitle

{\it Introduction.}---Crystal growth under diffusion-limited conditions often produces nonequilibrium morphologies that deviate from their equilibrium faceted shapes, such as dendritic~\cite{RevModPhys.52.1, ben1990formation} and hopper crystals~\cite{Amelinckx01031953, Sunagawa1999, FONTANA2011207, YIN2014131, Desarnaud2018, Pettit2019, https://doi.org/10.1002/adfm.201908108,Bollada2023}. Such morphologies have long been studied to uncover the underlying mechanisms of crystal growth. In particular, hopper crystals display a characteristic hollowed morphology, where the corners advance faster than the central regions of the crystal.
Although hopper crystals are generally attributed to the nonuniform solute diffusion field surrounding the growing facets~\cite{Sunagawa1999, Pimpinelli_Villain_1998, sunagawa2007crystals, https://doi.org/10.1002/adfm.201908108}, it remains an open question whether intrinsic material properties contribute to the occurrence of hopper growth.

The discovery of topological materials, such as topological insulators \cite{RevModPhys.82.3045, RevModPhys.83.1057},  topological crystalline insulators \cite{PhysRevLett.106.106802, hsieh2012topological, slager2013space}, higher-order topological insulators \cite{PhysRevLett.108.126807,PhysRevLett.110.046404,benalcazar2017quantized, PhysRevB.96.245115, PhysRevLett.119.246401, PhysRevLett.119.246402, schindler2018higher}, has shed new light on surface physics. 
The boundary states of these topological materials play crucial roles in determining surface energies and equilibrium crystal shapes~\cite{PhysRevLett.129.046802, PhysRevB.107.245148, Mondal2025}. 
We thus expect that topological phases will significantly influence nonequilibrium crystal growth, but this aspect remains largely unexplored.
Interestingly, candidate materials for higher-order topological phases hosting hinge states or corner charges, such as bismuth~\cite{schindler2018higherbismuth}, lead telluride~\cite{Robredo2019}, and sodium chloride~\cite{Watanabe2021}, often exhibit hopper-like crystal morphologies~\cite{FONTANA2011207, YIN2014131, Desarnaud2018, Pettit2019}. 
Yet, the role of higher-order topology in crystal growth remains elusive.

In this Letter, we show that higher-order topological phases give rise to a hollowed crystal morphology, characterized by preferential growth at the corners, through a previously unexplored mechanism that originates from the intrinsic electronic properties of the material.
We quantify the morphological difference between the two-dimensional (2D) higher-order topological and normal insulator phases using two fractal dimensions:~the overall fractal dimension $D_f$~\cite{Hausdorff1919, Manderlbrot_fractal_nature}, characterizing corner advancement, and the fractal dimension of coastlines $D_{f,c}$~\cite{doi:10.1126/science.156.3775.636, Manderlbrot_fractal_nature}, quantifying interfacial complexity.
When we compare crystals in the topological and normal insulator phases with the same $D_{f}$, we find that the former is in the hopper-like shape, while the latter is in the dendritic shape. Quantitatively, the higher-order topological phase exhibits a smaller $D_{f,c}$ than the normal insulator.

{\it In-gap states localized at corners.}---
We discuss in-gap states localized at the corner in a 2D higher-order topological insulator \cite{benalcazar2017quantized, PhysRevB.96.245115, PhysRevLett.119.246401, PhysRevLett.119.246402}. 
We begin with a simple 2D tight-binding model \cite{PhysRevLett.119.246402} on a square lattice with the lattice sites denoted by a lattice vector $\boldsymbol{R}=(x,y)$ with $x \in \mathbb{Z}$ and $y \in \mathbb{Z}$. 
At each lattice site, there exist four orbitals denoted by $\ket{\mu_z}\otimes \ket{\tau_z}$ ($\mu_z, \tau_z=\pm1$) with the spin degree of freedom denoted by $\ket{\sigma_z}$ ($\sigma_z=\pm1$). The Bloch Hamiltonian of this model is given by 
\begin{align}\label{eq:2dsoti}
	\mathcal{H}(\boldsymbol{k})= &m \tau_z-t\sum_{i=x,y}\cos k_i \tau_z  +v \tau_x  \biggl( \sum_{i=x,y} \sin k_i \sigma_i \biggr) \nonumber \\
	&+\Delta(\cos k_x - \cos k_y)\mu_y \tau_y,
\end{align}
where $\mu_i$, $\tau_i$, and $\sigma_i$  ($i=x,y,z$) are Pauli matrices with $m, t, v, \Delta \in \mathbb{R}$.
This model respects four-fold rotational symmetry $C_4 \mathcal{H}(\boldsymbol{k})C_4^{-1}=\mathcal{H}(\hat{C}_4 \boldsymbol{k})$ with $C_4 = \tau_{z}e^{-i\pi \sigma_z /4}$ and  $\hat{C}_4  \boldsymbol{k} =(-k_y, k_x)$.
In the following, we assume that the Fermi energy is zero, which means that four bands among the eight bands are occupied. Meanwhile, we choose the parameter values to satisfy $|m|<2t$ ($t>0$) so that this model exhibits a higher-order topological phase, or, more specifically, an obstructed atomic insulator (OAI) phase~\cite{bradlyn2017topological, po2017symmetry, benalcazar2017quantized, PhysRevB.96.245115, PhysRevB.97.035139, PhysRevB.99.245151, PhysRevLett.121.126402, PhysRevResearch.1.033074, PhysRevResearch.1.033074, PhysRevB.103.205123, kooi2021bulk}, which is characterized by the filling anomaly \cite{PhysRevB.99.245151, PhysRevResearch.1.033074, PhysRevB.102.165120, PhysRevB.103.205123, kooi2021bulk}~(see \SM~Sec.~I for the filling anomaly~\cite{sup1}). 

\begin{figure}
	\includegraphics[width=1.\columnwidth]{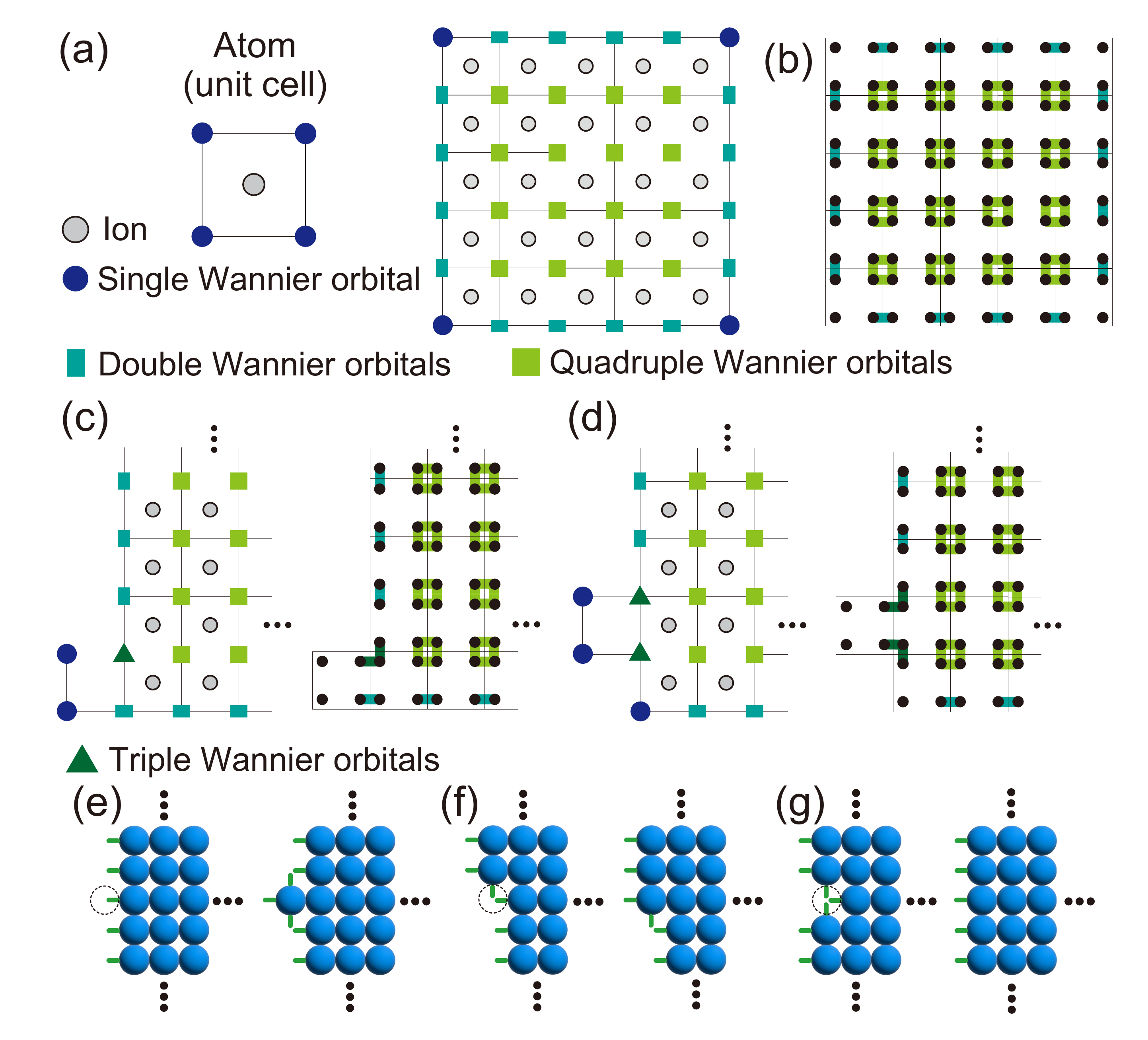}
	\caption{(a) The positions of ions and Wannier centers for the 2D model $\mathcal{H}(\boldsymbol{k})$. (b) A crystal with electronic orbitals located near the corners of the unit cell, which can be regarded as a covalent crystal between orbitals in the neighboring atoms. Black circles are the electronic orbitals, and the lines between the electrons indicate the covalent bonds. (c) A square-shaped crystal with an atom deposited at the corner. (d)  A square-shaped crystal with an atom deposited next to the corner. (e-g) The number of missing neighbors (the number of green segments) changes at the deposition of an atom onto the site at the dotted circle. }
	\label{fig:wc_corner}
\end{figure}

Within this phase, the positions of Wannier centers, \textit{i.e.}, the centers of the Wannier functions of occupied states, are the corners of the unit cells in the model given by Eq.~(\ref{eq:2dsoti})  [Fig.~\ref{fig:wc_corner}(a)] \cite{PhysRevLett.119.246402}. 
Each unit cell corresponds to one atom having an ion with charge $+4e$ located at the center of each unit cell and four electrons with charge $-e$. 
The OAI phase is characterized by this mismatch between the Wannier centers and the ionic positions, in contrast to a normal insulator where Wannier centers coincide with the ionic positions.
We consider this model in a finite-sized square, as shown in Fig.~\ref{fig:wc_corner}(a). 
Then, this system can be regarded as a covalent crystal with covalent bonds formed by pairs of the Wannier orbitals at the corners of the neighboring unit cells [Fig.~\ref{fig:wc_corner}(b)].
Here, as shown in Fig.~\ref{fig:wc_corner}(a), some Wannier orbitals from different unit cells meet together, and we classify the cases by the numbers (1, 2, 3, and 4) of the Wannier orbitals meeting together. We call them single, double, triple, and quadruple Wannier orbitals, respectively. 
We expect that the parity of this number indicates whether there are in-gap states.
If an even number of Wannier orbitals meet at the same point, the Wannier orbitals exhibit bonding-antibonding splitting, which lowers the occupied-state energy. Therefore, we expect that the double and quadruple Wannier orbitals lead to stable electronic states.  
On the other hand, single and triple Wannier orbitals have an electronic state that does not form pairs with other orbitals. 
This isolation results in the emergence of the in-gap states localized at the corners. 

As shown in Sec.~II of \SM~\cite{sup1}, our model $\mathcal{H}(\boldsymbol{k})$ is adiabatically connected to the direct sum of two copies of the prototypical Benalcazar-Bernevig-Hughes (BBH) model $\mathcal{H}_{\rm BBH}(\boldsymbol{k})$ \cite{benalcazar2017quantized}, which hosts corner states:
	$\mathcal{H}(\boldsymbol{k}) \rightarrow
	\mathcal{H}_{\rm BBH}(\boldsymbol{k}) \oplus  \mathcal{H}_{\rm BBH}(\boldsymbol{k}).$
In the BBH model, four orbitals are located near four corners of the unit cell, and the nearest-neighbor hopping between the orbitals exists, similar to Fig.~\ref{fig:wc_corner}(b).
The connection between our model $\mathcal{H(\boldsymbol{k})}$ and the BBH model supports our previously mentioned interpretation based on the Wannier orbitals.

{\it Energy changes in crystal growth.}---
We consider the number of in-gap states for a crystal shape where a single atom is attached to the corner of a square crystal [Fig.~\ref{fig:wc_corner}(c)].
In this geometry, two single Wannier orbitals and a triple Wannier orbital appear around the corner. 
Here, we introduce integers $n_s$ and $n_t$ representing the numbers of single Wannier orbitals and triple Wannier orbitals, respectively.
Triple Wannier orbitals correspond to the mixture of three Wannier functions, leading to three states: a bonding state, an antibonding state, both in the bulk bands, and a state in the gap.
Thus, $n_s + n_t$ is equal to the number of in-gap states in Fig.~\ref{fig:wc_corner}(c).
By considering that each of the other three corners of the original square crystal has a single Wannier orbital, we get $n_s = 5$ and $n_t=1$ for this geometry, resulting in the emergence of six in-gap states. 
Next, we consider a square crystal with an atom attached next to the corner, as shown in Fig.~\ref{fig:wc_corner}(d). We find that this system with this geometry leads to $n_s = 6$ single Wannier orbitals and $n_t=2$ triple Wannier orbitals, and therefore eight in-gap states appear. 
Thus, we expect that the energy change upon the deposition of an atom onto the corner [Fig.~\ref{fig:wc_corner}(c)] is lower than that onto the edges away from the corners [Fig.~\ref{fig:wc_corner}(d)]. 

We introduce the energy change $\Delta E_{n} = E(n) - E(n-1)$, where $E(n)$ is the sum of electronic energies of all the occupied states for the system with $n$ atoms. Here, in order to highlight the feature of the OAI, we do not consider the electrostatic energy of the nuclei and electron-electron interactions for simplicity.
We find that $\Delta E_{n}$ can be approximated as
\begin{equation}\label{eq:energy_change_approx}
	\Delta E_{n} \simeq  \Delta n_b e_b + \mu_s +\Delta {n}_{s} e_{s}+\Delta {n}_{t} e_{t},
\end{equation}
from the numerical result~[Fig.~\ref{fig:energy_change}] in End Matter. 
Here, $\mu_s$ is the chemical potential of the solid phase, $\Delta n_s$ ($\Delta n_t$) is the change in $n_s$ ($n_t$) during the deposition of an atom, and $\Delta n_b$ indicates the changes of the number of missing neighbors (or dangling bonds) in the deposition of an atom, which we explain later.  
The parameters $e_s$, $e_t$, and $e_b$ are fitting parameters. Then $e_s$ ($e_t$) represents the energies of an in-gap state induced by the single (triple) Wannier orbitals, 
and $e_b$ represents the energy associated with undercoordination. 
Here, we explain the definition of the number of missing neighbors in our model. 
If one adjacent site of an atom is not occupied by another atom, we regard the adjacent site to be a missing neighbor [Fig.~\ref{fig:wc_corner}(e-g)] \cite{saito1996book,jeong1999steps,gilmer_simulation_2003,huang2008statistical} because our model has hoppings along the $x$ and $y$ directions. 
Thus, the change in the number of missing neighbors $\Delta n_b$ upon deposition of an atom at $(x,y)$ depends on how many of its four nearest-neighbor sites $(x\pm1,y)$, $(x,y\pm1)$ are already occupied: 
$\Delta n_b=2$ when one is occupied [Fig.~\ref{fig:wc_corner}(e)], $\Delta n_b=0$ when two are occupied [Fig.~\ref{fig:wc_corner}(f)], and $\Delta n_b=-2$ when three are occupied [Fig.~\ref{fig:wc_corner}(g)].
The first and second terms in Eq.~(\ref{eq:energy_change_approx}) are typically included in the theories of growth of covalent crystals in the previous works \cite{gilmer_simulation_2003, PhysRevB.29.328, PhysRevA.38.2447, PhysRevA.40.3408}.
In contrast, the third and fourth terms are newly added terms characteristic of the OAI (see \SM~Sec.~III for the formulae of $\Delta n_s$ and $\Delta n_t$~\cite{sup1}). 

{\it Crystal growth induced by corner states.}---
We consider a simple Monte Carlo model of a growing crystal in a diffusion field \cite{PhysRevA.40.3408}.
This model is defined on the square lattice with the lattice vector $\boldsymbol{R}=(x,y)$ and $x, y \in \mathbb{Z}$, and the crystal is described by the Hamiltonian $\mathcal{H(\boldsymbol{k})}$.
Because of four-fold rotational symmetry of the Hamiltonian $\mathcal{H(\boldsymbol{k})}$, we consider the first quadrant with $x \geq 0$ and $y \geq 0$ of the lattice for simplicity and  impose the periodic boundary condition connecting the positive $x$-axis and the positive $y$-axis. 
In this model, each lattice site at $\boldsymbol{R}=(x,y)$ ($x\geq 0$, $y\geq 0$) can be in one of three states: occupied by a gas atom, occupied by a solid atom, or vacant.
The gas atoms move randomly to a nearest-neighbor site. If a gas atom moves to a site adjacent to a solid atom, it becomes a solid atom with a probability $W=(1+\exp(\Delta E/k_{\rm B}T))^{-1}$. Here, we define the energy change associated with the deposition of a gas atom as $\Delta E:=\Delta E_n-\mu_g$, where $\Delta E_n$ is given by Eq.~\eqref{eq:energy_change_approx}, and $\mu_g$ is the chemical potential of the gas phase.
This energy change is written by
\begin{align}\label{eq:delta_E}
    \Delta E \simeq  \Delta n_b e_b +\Delta {n}_{s} e_{s}+\Delta {n}_{t} e_{t} +\Delta \mu - k_{\rm B}T \ln n_g,
\end{align}
where $\Delta \mu := \mu_s - \mu_{g, c}$ ($<0$) with $\mu_{g,c}$ being the energy per gas atom, and $n_g$ is the gas density (see Sec.~IV of \SM~for the details on the Monte Carlo simulation of crystal growth~\cite{sup1}).

\begin{figure}
	\includegraphics[width=1.\columnwidth]{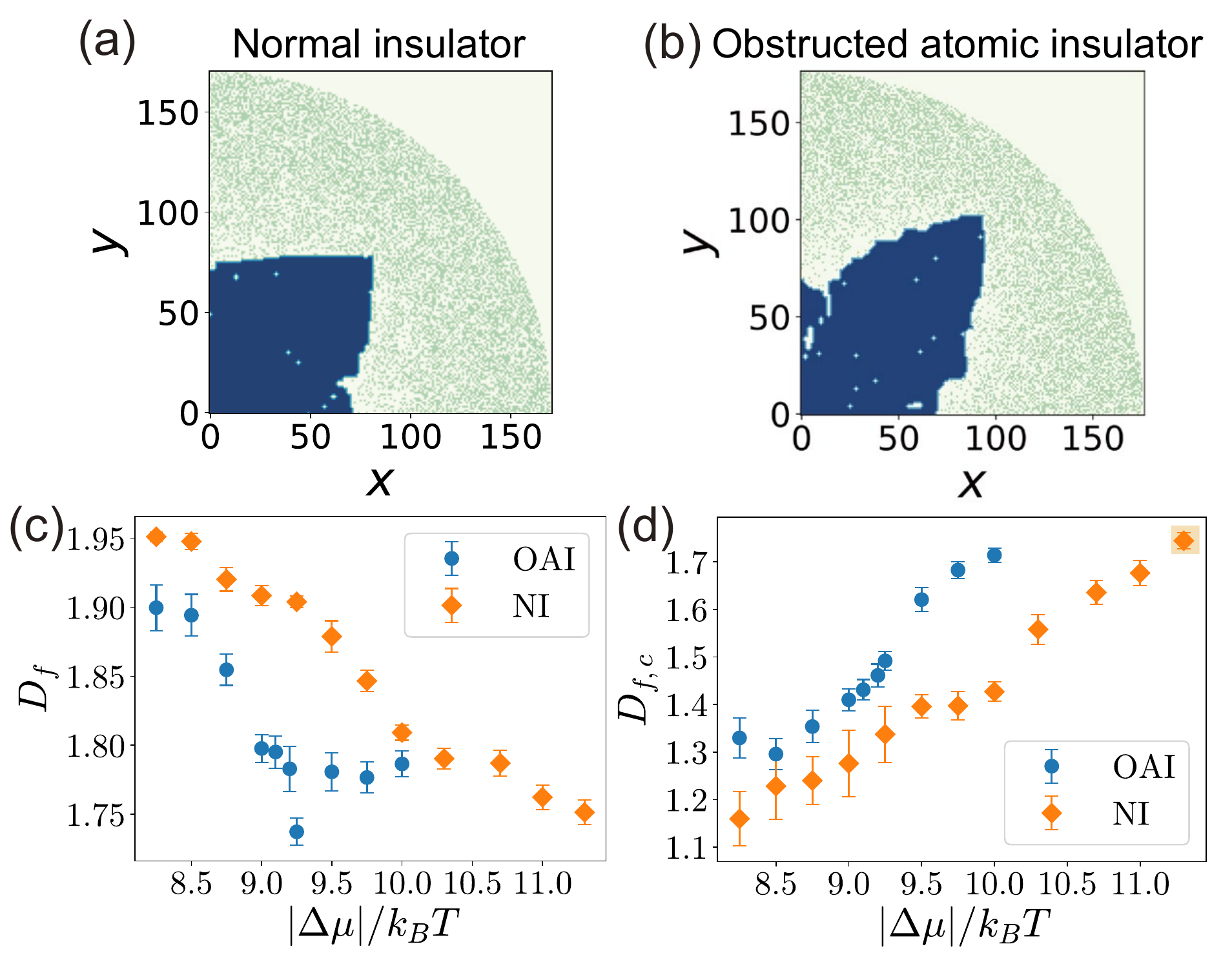}
	\caption{(a,b)~Crystal shapes obtained from the simulations. 
    The blue, light blue, and light green sites indicate the solid atoms, interfacial solid atoms, and gas atoms, respectively. The parameters in Eq.~(\ref{eq:energy_change_approx}) are set to (i) $e_b=0.6439+0.2413+0.1488$, $e_s = 0$, $e_t = 0$ in (a) corresponding to the normal insulator and (ii) $e_b=0.6439$, $e_s = 0.2413$, $e_t = 0.1488$ in (b) corresponding to the OAI, which are the values obtained by the fitting to the numerical result~[Fig.~\ref{fig:energy_change}] in End Matter. The crystal shapes are obtained with $t=1500$ and $\Delta \mu/k_{\rm B}T = -8.75$. We choose $k_{\rm B}T=0.16$ in both the (i) and (ii) cases.
    (c)~Fractal dimensions $D_f$. (d)~The fractal dimension of coastlines $D_{f,c}$. Here, ``NI'' and ``OAI'' indicate the parameters (i) for the normal insulator and (ii) for the OAI, respectively.
    The error bars indicate the standard error of the mean of six simulations for each value of $\Delta \mu/k_{\rm B}T$. To reduce statistical uncertainty, we increase the number of Monte Carlo samples to twenty at $|\Delta \mu|/k_BT \geq 10$ for the parameters (i) and at $|\Delta \mu|/k_BT \geq 9$ for the parameters (ii). The fits for the fractal dimensions are performed only for data points with $N_{s,i}>100$.
}
	\label{fig:growth_shape}
\end{figure}

We perform the simulation with two types of parameters:~(i) a normal insulator case:~$(e_b,e_s,e_t)=(e^{(\mathrm{n})}_b, 0, 0)$, and (ii) an OAI case:~$(e_b,e_s,e_t)=(e_b^{(\mathrm{OAI})}, e_s^{(\mathrm{OAI})}, e_t^{(\mathrm{OAI})})$.
Throughout this work, we take the parameter values $e_b^{(\mathrm{OAI})}=0.6439$, $e_s^{(\mathrm{OAI})} =  0.2413$, and $e_t^{(\mathrm{OAI})}=0.1488$, which are the values obtained by the fitting to the numerical result~[Fig.~\ref{fig:energy_change}] in End Matter.
We put the value of $e^{(\mathrm{n})}_b$ to be $e^{(\mathrm{n})}_b=e_b^{(\mathrm{OAI})} + e_s^{(\mathrm{OAI})} + e_t^{(\mathrm{OAI})}$ so that $e_b + e_s + e_t$ is identical between (i) and (ii) for comparison between (i) and (ii) (see Sec.~V of Supplemental Material for the case of $e^{(\mathrm{n})}_b=e_b^{(\mathrm{OAI})}$~\cite{sup1}).
In the prototypical deposition processes onto the surface in Fig.~\ref{fig:wc_corner}(e-g), the energy changes are given by $\Delta E_{n} = 2(e_b+ e_s + e_t)$ in Fig.~\ref{fig:wc_corner}(e), $\Delta E_{n} = 0$ in Fig.~\ref{fig:wc_corner}(f), and $\Delta E_{n} = -2(e_b+ e_s + e_t)$ in Fig.~\ref{fig:wc_corner}(g).
Thus, by setting the value of $e_b+ e_s + e_t$ to be identical for both (i) and (ii), we can compare these cases under the condition of identical $\Delta E_{n}$ for the deposition onto the surface in the three cases~[Fig.~\ref{fig:wc_corner}(e-f)].
Figures~\ref{fig:growth_shape}(a) and  \ref{fig:growth_shape}(b) show the crystal shapes obtained in the simulation with (i) the normal insulator phase [Fig.~\ref{fig:growth_shape}(a)] and (ii) the OAI phase  [Fig.~\ref{fig:growth_shape}(b)].
From these results, we find the corner grows faster than the edges in the presence of the topological in-gap states, favoring the growth in the $\langle  11 \rangle$ direction.

\begin{figure}
	\includegraphics[width=1.\columnwidth]{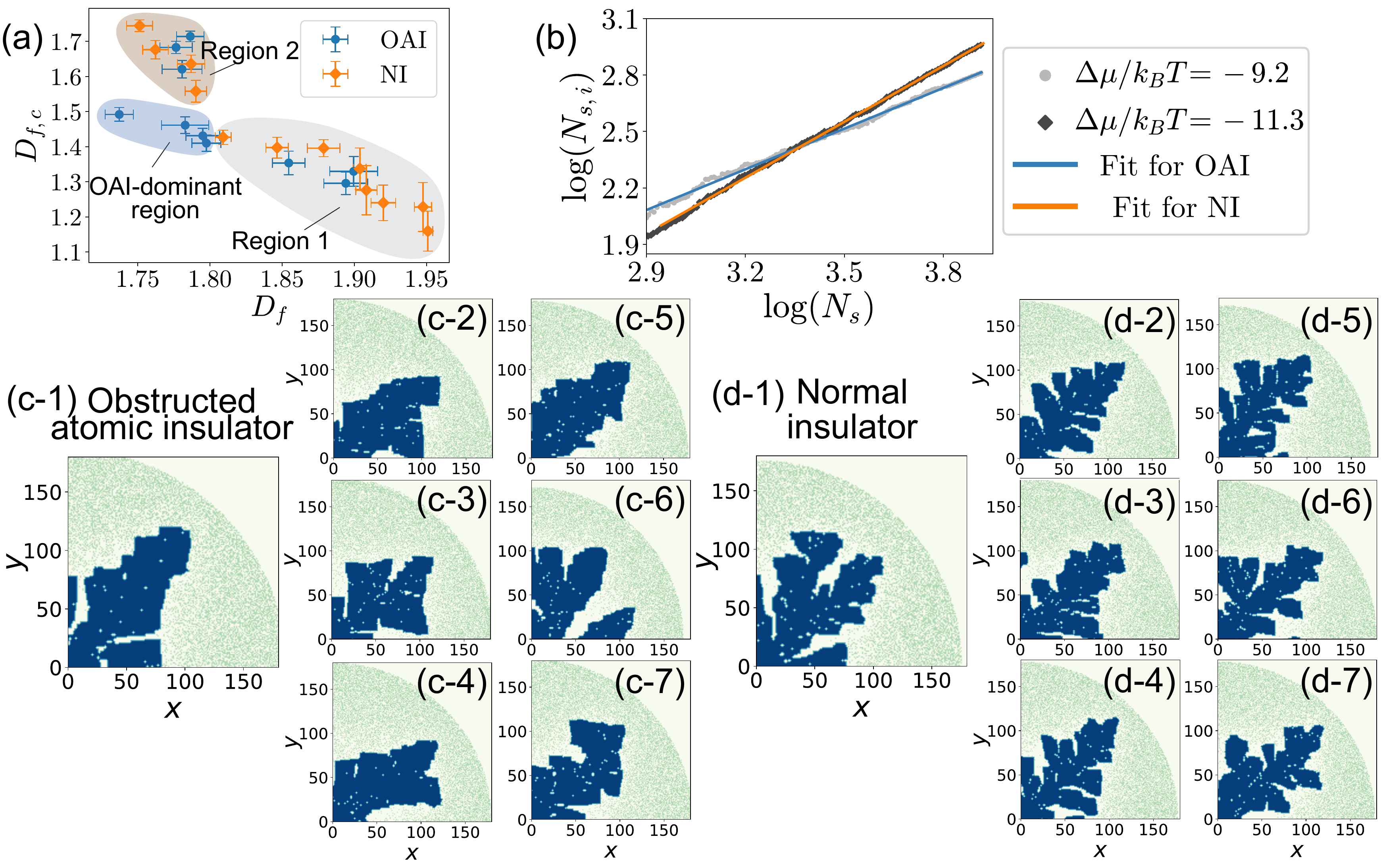}
	\caption{(a)~Fractal dimensions $D_f$ versus fractal dimensions of coastlines $D_{f,c}$.  Here, ``OAI'' and ``NI'' (normal insulator) indicate the parameters (i) and (ii) in Fig.~\ref{fig:growth_shape}, respectively. The error bars indicate the same standard errors as those in Fig.~\ref{fig:growth_shape}.
    (b)~The relationship between the logarithm of $N_s$ and the logarithm of $N_{s,i}$, where the logarithm base is 10. The fits are performed only for data points with $N_{s,i}>100$. (c,d) The obtained crystal shapes for (c) the parameters (i) in Fig.~\ref{fig:growth_shape} with $\Delta \mu/k_{\rm B}T=-9.2$ and (d) the parameters (ii) in Fig.~\ref{fig:growth_shape} with $\Delta \mu/k_{\rm B}T=-11.3$, with $8000\leq N_{s} \leq 8010$. (c-1) and (d-1) correspond to the results in (b). The corresponding times for (c-1)-(c-7) and (d-1)-(d-7) are summarized in Table~\ref{tab:df_dfc_list}.
 }
	\label{fig:fractal_characters}
\end{figure}

To characterize the crystal shapes, we introduce a fractal dimension for 2D objects, which indicates how much space is filled by the object \cite{Manderlbrot_fractal_nature}.
The mass of the crystal $N_s$ scales with its characteristic length $e.g.$, a radius of the gyration $R_g$ as follows:
\begin{align}\label{eq:2dfractal}
	N_s \propto R_g^{D_f},
\end{align}
where $D_f$ is the fractal dimension. The radius of gyration $R_g$ is given as 
$
	R_g = \sqrt{ \sum_{i=1}^{N_s} {r_i^2}/{N_s}},
$
where $r_i$ is the distance between the origin and the position of the $i$-th solid atom, and $i$ runs over all the solid atoms. 
The fractal dimension $D_f$ takes values between 1 and 2, and a 2D disk has the fractal dimension $D_f=2$. 
Figure~\ref{fig:growth_shape}(c) shows that the fractal dimension $D_f$ in the OAI phase is smaller than that in the normal insulator phase for the same value of $\Delta \mu$ $(<0)$.
When the magnitude $|\Delta \mu|$ increases, the crystal grows faster, and the fractal dimension $D_f$ becomes lower both in the normal insulator and in the OAI phases.
This behavior arises because the Mullins-Sekerka interface instability, driven by diffusion in the surrounding field, occurs when the growth rate becomes sufficiently high \cite{Mullins1963}. 
As discussed in End Matter, both the OAI and the normal insulator exhibit comparable growth rates at similar $D_f$. Thus, we compare the crystal morphologies of these two phases based on similar $D_f$ (see Sec.~VI of Supplemental Material for a comparison between these two phases at the same $\Delta \mu$). 

\begin{table*}
	\centering
	\caption{The energy changes $\Delta E/(k_{\rm B}T)$ in the  representative examples of the growth process in Fig.~\ref{fig:deposition_position}.
    Here,  ``NI'' indicates the normal insulator.
    In the columns ``Fig.~\ref{fig:fractal_characters}(c)'' and ``Fig.~\ref{fig:fractal_characters}(d)'', we calculate the energy changes $\Delta E/(k_{\rm B}T)$ with the same parameters in Figs.~\ref{fig:fractal_characters}(c) and  \ref{fig:fractal_characters}(d). 
    Here we choose the gas density $n_g=0.5$. 
    }
	\label{tab:energy_change}
     \begin{tabular}{ccccc} \hline \hline
    Position & OAI & NI & Fig.~\ref{fig:fractal_characters}(c) (OAI)~ & Fig.~\ref{fig:fractal_characters}(d) (NI)  \\ \hline 
    ~$l=1$ & $(2e^{(n)}_b + e^{(n)}_s + e^{(n)}_t+\Delta \mu)/(k_{\rm B}T)-\ln n_g$~ & ~$(2e^{(t)}_b+\Delta \mu)/(k_{\rm B}T)-\ln n_g$~  & 1.98 & 2.32
    \\
    $l=2$ & $(2(e^{(n)}_b + e^{(n)}_s + e^{(n)}_t)+\Delta \mu)/(k_{\rm B}T)-\ln n_g$ & $(2e^{(t)}_b+\Delta \mu)/(k_{\rm B}T)-\ln n_g$  & 4.42 & 2.32 \\ 
        $l=3$ & $(-e^{(n)}_s - e^{(n)}_t+\Delta \mu)/(k_{\rm B}T)-\ln n_g$ & $\Delta \mu/(k_{\rm B}T)-\ln n_g$  & $-11.8$ & $-10.6$  \\
    \hline \hline
  \end{tabular}
\end{table*}

To clarify the difference between the two phases, we introduce the fractal dimension of coastlines $D_{f,c}$ \cite{doi:10.1126/science.156.3775.636, Manderlbrot_fractal_nature} to evaluate how the space is filled by the curve [Fig.~\ref{fig:growth_shape}(d)]. 
The number of solid atoms at the interface $N_{s,i}$ scales with the radius of gyration $R_g$  
\begin{align}\label{eq:coastline_fractal}
	N_{s,i} \propto R_g^{D_{f,c}}.
\end{align}
$D_{f,c}$ takes values from 1 to 2, and an ordinary curve has $D_{f,c} \approx 1$, while a space-filling curve, such as the Hilbert curve \cite{Hilbert1891459} exhibits $D_{f,c} \approx 2$.
Figure~\ref{fig:fractal_characters}(a) shows the $D_f$-$D_{f,c}$ plot by changing $|\Delta \mu|$ for the two phases. In the normal insulator phase (orange symbols) at low growth rates~($|\Delta \mu|<10.3k_{\rm B}T$), which we refer to as Region 1, the crystal shape is governed by the equilibrium crystal shape, which is a square here. In contrast, at high growth rates~($|\Delta \mu|\geq 10.3k_{\rm B}T$), which we refer to as Region 2, it exhibits a dendritic morphology~(see End Matter for the details on the growth at high growth rates). Thus, in the normal insulator, by increasing $|\Delta \mu|$, the growth rate becomes faster, resulting in a monotonic decrease of $D_f$ and a monotonic increase of $D_{f,c}$. In contrast, in the OAI, in between Region 1 ($|\Delta \mu|<9k_{\rm B}T$) and Region 2 ($|\Delta \mu|>9.25k_{\rm B}T$), a new region, which we call the OAI-dominant region, appears at the intermediate growth rate~($9k_{\rm B}T\leq|\Delta \mu|\leq 9.25k_{\rm B}T$)~[the data within the blue-shaded region in Fig.~\ref{fig:fractal_characters}(a)]. As we increase $|\Delta \mu|$, $D_f$ first decreases, and the crystal shape changes from Region 1 to the OAI-dominant region, which is characterized by a smaller value of $D_{f,c}$ than Region 2. Then $D_f$ and $D_{f,c}$ increase to enter Region 2. As shown in Fig.~\ref{fig:growth_shape}(c), $D_f$ shows a non-monotonic behavior. Thus, the OAI-dominant region is unique to the OAI and is inaccessible to the normal insulator phase. 

\begin{figure}
	\includegraphics[width=1.\columnwidth]{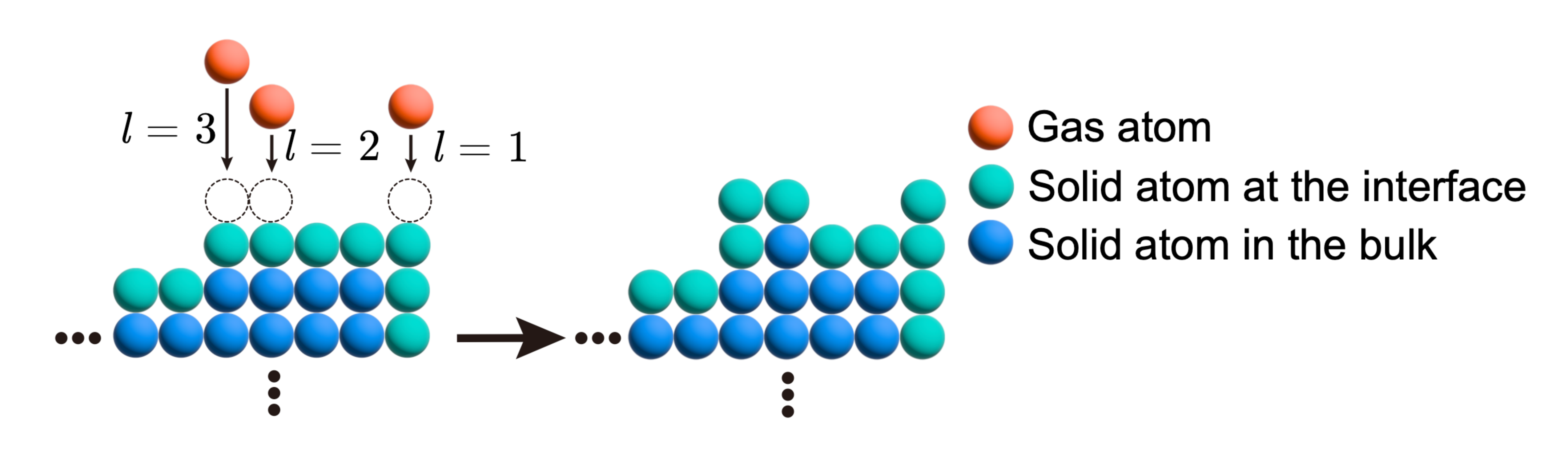}
	\caption{Representative examples of the growth process that increase the number of atoms at the interface $N_{s,i}$.
}
	\label{fig:deposition_position}
\end{figure}

By combining Eq.~(\ref{eq:2dfractal}) and Eq.~(\ref{eq:coastline_fractal}), we obtain the following relation between the perimeter ($N_{s,i}$) and the area ($N_{s}$) in the crystal morphology \cite{Manderlbrot_fractal_nature, mandelbrot1984fractal, IMRE2006443, FLORIO2019551}:
\begin{align}\label{eq:perimeter_area_relation}
N_{s,i} \propto N_{s}^{D_{f,c}/D_{f}}.
\end{align}
Figure~\ref{fig:fractal_characters}(b) shows the perimeter-area relationship in the simulation. 
We perform a linear fit to the log-log plots of $N_{s}$ and $N_{s,i}$ and find that the slopes ($=D_{f,c}/D_{f}$) for the normal insulator parameters are steeper than the OAI cases.
This behavior arises because the normal insulator phase possesses a larger $D_{f,c}$ than the OAI phase, by comparing the crystals in both phases with similar values of $D_f$.  
Thus, the difference in the crystal growth between the normal insulator and OAI phases emerges in the perimeter-area relationship.

Such a difference between the normal insulator and OAI phases manifests in the crystal shape.
In Fig.~\ref{fig:fractal_characters}(c), the seven panels show the simulation results with the same parameters for the OAI phase, and the results for the normal insulator phase are shown in Fig.~\ref{fig:fractal_characters}(d). 
In the normal insulator phase, the crystal shapes are dendritic, while in the OAI phase, they exhibit hollowed morphologies characterized by the preferential growth at the corners with a relatively smooth interface.
The difference between the normal insulator and OAI phases is quantified by $D_{f,c}$.
While the fractal dimensions $D_f$ are $1.78$ in Fig.~\ref{fig:fractal_characters}(c-1) and $1.77$ in Fig.~\ref{fig:fractal_characters}(d-1) (calculated for the single realizations) and are of similar size, the fractal dimensions of coastlines $D_{f,c}$ are different, \textit{i.e.}, $1.29$ in Fig.~\ref{fig:fractal_characters}(c-1) and $1.75$ in Fig.~\ref{fig:fractal_characters}(d-1). The values of $D_f$ and $D_{f,c}$ for panels (c-2)–(c-7) and (d-2)–(d-7) in Fig.~\ref{fig:fractal_characters} are summarized in Table~\ref{tab:df_dfc_list} in End Matter. 
Namely, when we compare the crystal shapes for the OAI and normal insulator phases having the same fractal dimension $D_{f}$ $(<1.8)$, which means that the growth speed at the corners is of a similar order, the surface in the normal insulator phase has large $D_{f,c}$, \textit{i.e.}, it is rougher compared to that in the OAI phase.
Thus, the OAI phase exhibits the crystal shape with developed corners and a relatively smooth interface.
A hopper crystal is defined as a form bounded by facets with a stepwise depressed center~\cite{Sunagawa1999, sunagawa2007crystals}. The morphologies in Fig.~\ref{fig:fractal_characters}(c) share key geometric features with a hopper crystal: they possess a depressed center bounded by a relatively smooth interface. Thus, we refer to this crystal morphology as ``hopper-like.''

When the growth along the $\langle 11\rangle$ direction is preferred over the growth along the $\langle10\rangle$ and $\langle01\rangle$ directions, the fractal dimension $D_{f}$ decreases. Thus, the deposition at the corner [$l=1$ in Fig.~\ref{fig:deposition_position}] is the primary source of the decrease in the fractal dimension $D_{f}$. 
On the other hand, the fractal dimension of coastlines $D_{f,c}$ is governed by various processes, such as the deposition at the corner [$l=1$ in Fig.~\ref{fig:deposition_position}] as well as that at the edge [$l=2$ in Fig.~\ref{fig:deposition_position}].
The energy changes in various depositions at $l$ $(=1,2,3)$ are summarized in Table~\ref{tab:energy_change}, where the deposition at $l=3$ occurs after $l=2$.
Here we compare the crystals with a similar value of $D_{f}$ in Figs.~\ref{fig:fractal_characters}(c) and \ref{fig:fractal_characters}(d). Then, in agreement with our expectation, the energy changes for $l=1$ have similar values between the OAI and normal insulator cases, as shown in Table~\ref{tab:energy_change}. On the other hand, in the energy changes for $l=2$, the normal insulator phase has a much smaller value than the OAI phase, and the deposition at $l=2$ occurs more often in the normal insulator phase. Once an atom is deposited at $l=2$, the deposition at $l=3$ occurs immediately because the associated energy change $\Delta E$ is strongly negative.
Thus, to summarize, in the normal insulator phase, atoms can be deposited at the edge more easily than in the OAI phase, and such an adsorbed atom works as a nucleation site for crystal growth, leading to the rougher crystal shape with a large $D_{f,c}$. We illustrate the typical growth processes in Fig.~\ref{fig:growth_process} in End Matter.

{\it Discussion.}---In summary, we have theoretically demonstrated in the 2D higher-order topological insulator (\textit{i.e.}, the OAI) that the crystal will take the hopper-like shape, in contrast to the dendritic shape in the normal insulator phase, in the case of relatively rapid crystal growth. It is characterized by the smaller value of $D_{f,c}$ for the crystal in the topological phase, compared to the crystal in the normal insulator phase with a similar value of $D_{f}$. This may explain the famous hopper crystal shapes in bismuth, lead telluride, and sodium chloride, since they are candidate materials hosting higher-order topological phases~\cite{schindler2018higherbismuth, Robredo2019, Watanabe2021}.
While we concentrate on 2D crystals in this work, our theory can be extended directly to three-dimensional crystals, since the analysis of Wannier centers and the definitions of the fractal dimensions remain valid in three-dimensional crystals. 
Our theory assumes that the materials are covalent crystals and focuses on the OAI. 
Because our theory is based on the topological nature of materials, our findings are universally applicable to covalent crystals classified as OAIs. Such candidate materials are SiAs, GeP, and ${\rm BC}_3$, which are reported as OAIs in Ref.~\cite{Yuanfeng2024}.

\begin{acknowledgments}
\textit{Acknowledgments.}---We thank Akira Furusaki and Makio Uwaha for fruitful discussions. This work was supported by Japan Society for the Promotion of Science (JSPS) KAKENHI Grants No.~JP22K18687, No.~JP22H00108, No.~JP24H02231, and No.~JP24K22868, and by MEXT Initiative to Establish Next-Generation Novel Integrated Circuits Centers (X-NICS) Grant No. JPJ011438. Y.T. is supported by RIKEN Special Postdoctoral Researchers Program and by JST CREST Grant No.~JPMJCR19T2.
\end{acknowledgments}

%

\clearpage

\section{End Matter}
\begin{figure}[b]
	\includegraphics[width=1.\columnwidth]{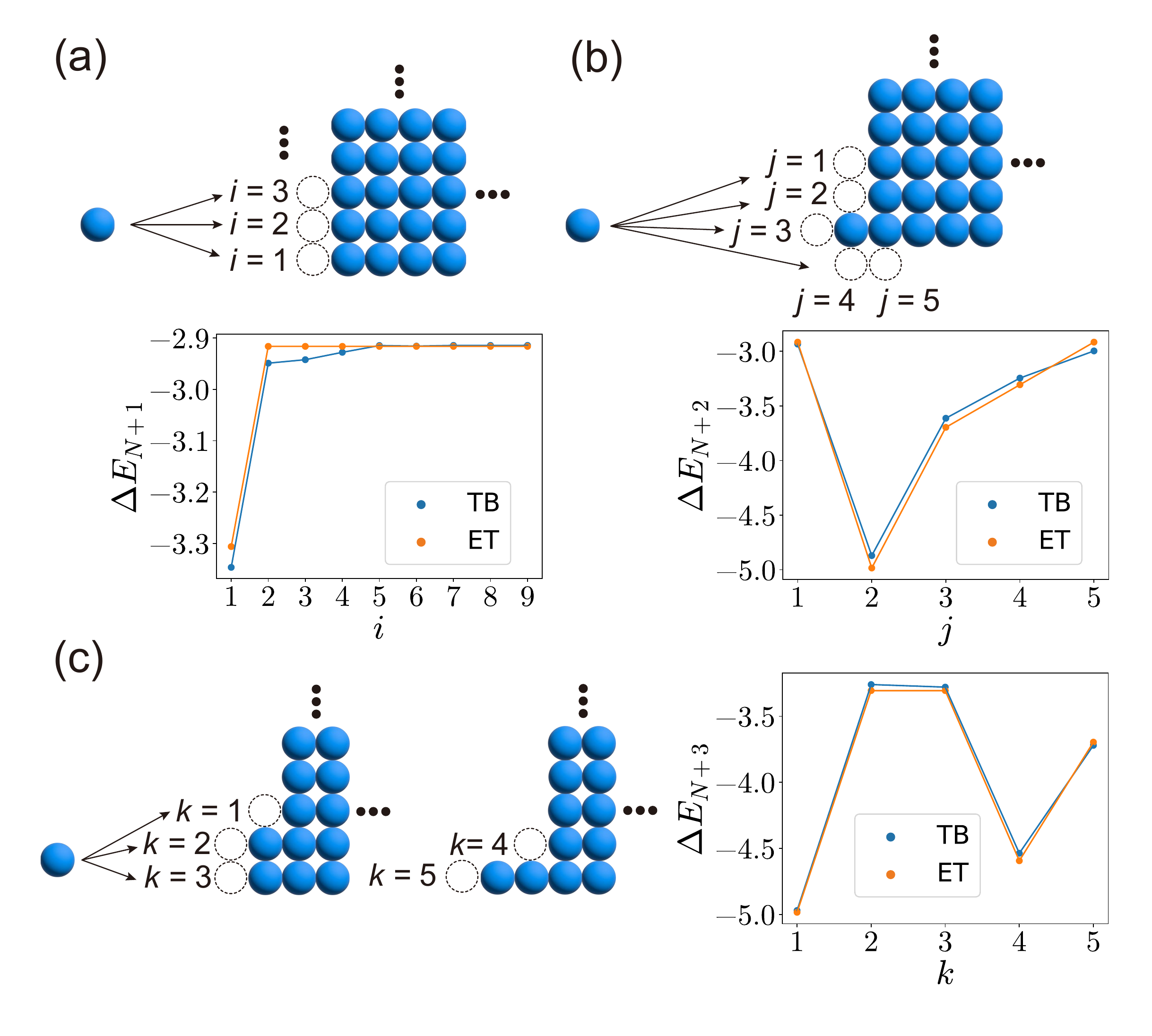}
	\caption{(a-c)~The energy changes upon the deposition of an atom onto crystals with
    (a) $N$ atoms (b) $N+1$ atoms, and (c) $N+2$ atoms. The positions of the deposition are labeled with $i$ in (a), $j$ in (b), and $k$ in (c). TB and ET indicate the calculations from the tight-binding model $\mathcal{H}(\boldsymbol{k})$ and those from the effective theory given in Eq.~(\ref{eq:energy_change_approx}), respectively. We choose the parameters $t=1$ and $m=v=\Delta = 0.5$ with the system with a square shape of $30\times 30$. The parameters of the approximation in Eq.~(\ref{eq:energy_change_approx}) are set to $\mu_s = -4.984$, $e_b=0.6439$, $e_s  =  0.2413$, and $e_t = 0.1488$.  
    }
	\label{fig:energy_change}
\end{figure}

{\it Numerical result of the energy change.}---Here, we calculate the change in the total energy of our model $\mathcal{H}(\boldsymbol{k})$ when one atom is adsorbed on the crystal surface. 
We begin with a square-shaped crystal with $N$ atoms.
The energy change is given by $\Delta E_{N+1} = E(N+1) - E(N)$, where $E(N)$ is the sum of electronic energies of all the occupied states for the system with $N$ atoms. 
Figure \ref{fig:energy_change}(a) shows the dependence of $\Delta E_{N+1}$ on the position of the adsorbed atom.
The result indicates that the energy change associated with the deposition of the atom at the corner ($i=1$) is lower than that on the edges ($i=2,3,\cdots$).
This result is consistent with our expectation based on the changes in $n_s$ and $n_t$. 
Furthermore, we also calculate $\Delta E_{N+2} = E(N+2) - E(N+1)$ [Fig.~\ref{fig:energy_change}(b)] and $\Delta E_{N+3} = E(N+3) - E(N+2)$ [Fig.~\ref{fig:energy_change}(c)] for various geometries. 
From the numerical result, we find that $\Delta E_{n}$ ($n=N+1, N+2, \cdots$) is well described by Eq.~\eqref{eq:energy_change_approx}.

\begin{figure}
	\includegraphics[width=1.\columnwidth]{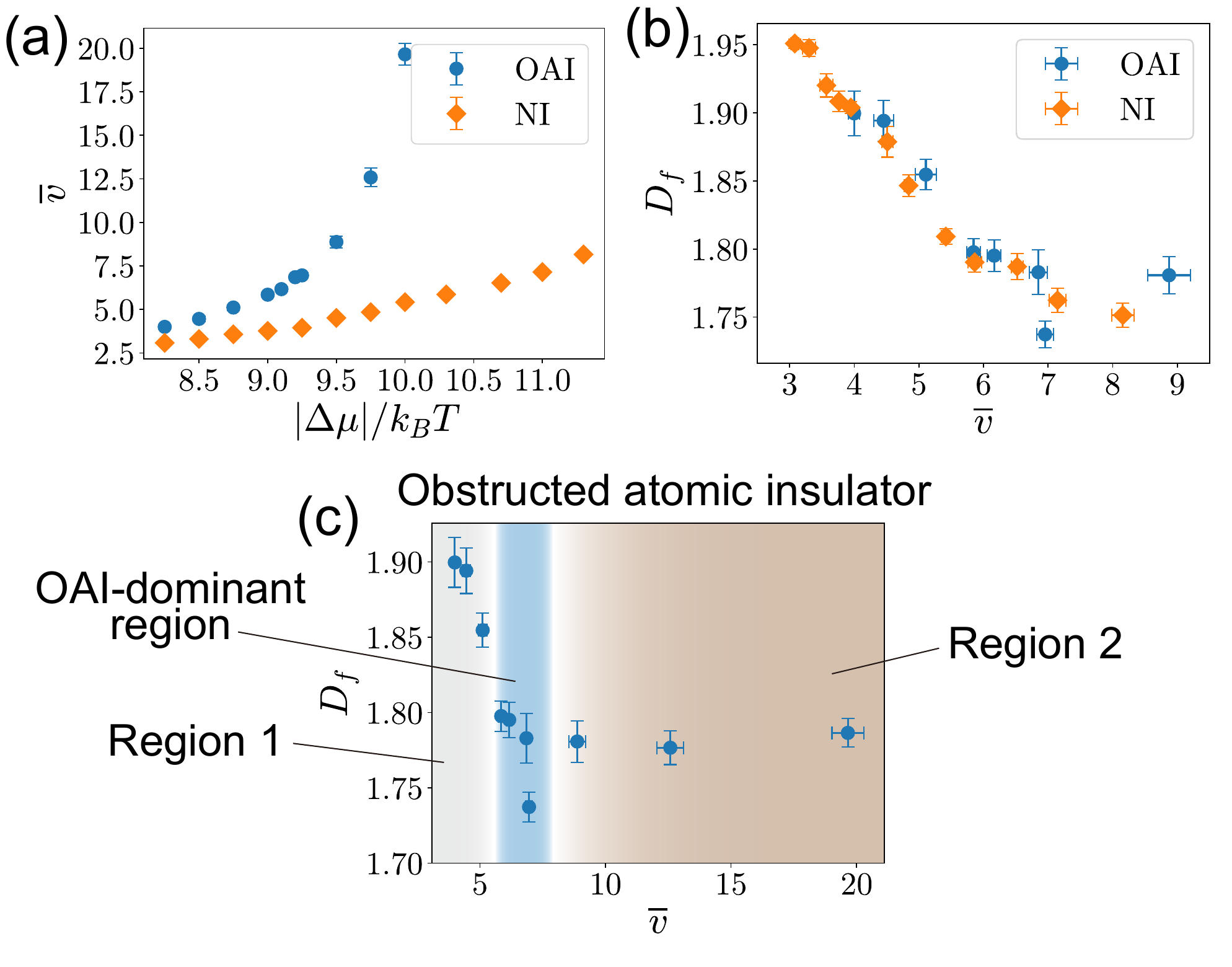}
	\caption{(a)~Growth rate $\bar{v}$ versus $|\Delta \mu|/k_{\rm B}T$. Here, ``OAI'' and ``NI'' (normal insulator) indicate the parameters (i) and (ii) in Fig.~\ref{fig:growth_shape}, respectively.
    (b)~Fractal dimension $D_{f}$ versus the growth rate $\bar{v}$. (c)~Three regions of the growth rate $\bar{v}$ for the OAI.
    The error bars indicate the standard error of the mean of six simulations for each value of $\Delta \mu/k_{\rm B}T$. To reduce statistical uncertainty, we increase the number of Monte Carlo samples to twenty at $|\Delta \mu|/k_BT \geq 10$ for the parameters (i) and at $|\Delta \mu|/k_BT \geq 9$ for the parameters (ii). The fits for the fractal dimension $D_f$ are performed only for data points where the number of solid atoms at the interface $N_{s,i}$ satisfies $N_{s,i}>100$.
    }
	\label{fig:growth_rate}
\end{figure}

{\it Growth rates.}---In this section, we discuss the growth rate in our simulation. 
We define the growth rate as
\begin{align}
    \bar{v}:=\frac{N_s - N_s^0}{t}, 
\end{align}
where $N_s$ and $N_s^0$ are the numbers of solid atoms at the end and start of the simulation, respectively, and $t$ is the time at the end of the simulation. Figure~\ref{fig:growth_rate}(a) shows that, even at the same value of $|\Delta \mu|/k_{\rm B}T$, the growth rate $\bar{v}$ of the OAI differs significantly from that of the normal insulator. On the other hand, both the OAI and the normal insulator exhibit comparable fractal dimension $D_f$ at similar growth rates $\bar{v}$~[Fig.~\ref{fig:growth_rate}(b)]. Consequently, comparing crystal morphologies with similar $D_f$ is equivalent to comparing them at similar growth rates. Therefore, we compare the results of these two phases based on similar values of $D_f$ in the main text.

Furthermore, to identify the growth rate $\bar{v}$ at which the OAI exhibits the crystal morphologies distinct from those of the normal insulator, we categorize the growth rate $\bar{v}$ into three regions: Region 1, Region 2, and the OAI-dominant region in Fig.~\ref{fig:growth_rate}(c). The growth rates of the OAI for $|\Delta \mu|<9k_{\rm B}T$ and $|\Delta \mu|>9.25k_{\rm B}T$ fall into Region 1 and Region 2, respectively. In these regions, the crystal morphologies of the OAI are indistinguishable from those of the normal insulator. In contrast, the growth rate in the intermediate regime ($9k_{\rm B}T\leq|\Delta \mu|\leq 9.25k_{\rm B}T$) falls into the OAI-dominant region in Fig.~\ref{fig:growth_rate}(c), where the OAI exhibits the characteristic morphology absent in the normal insulator, as discussed in the main text. 

\begin{figure}
	\includegraphics[width=1.\columnwidth]{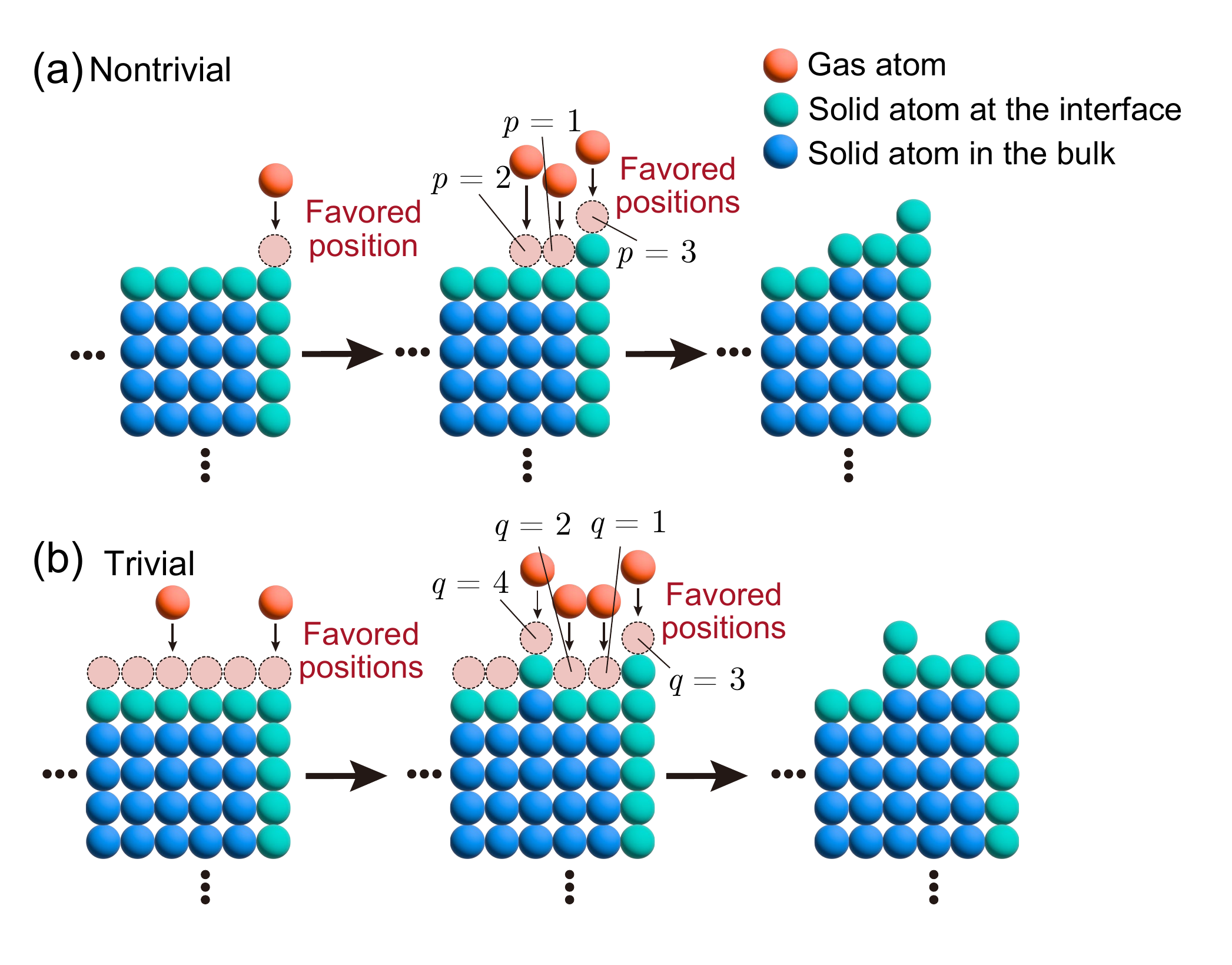}
	\caption{Representative examples of the growth process in (a) the OAI and (b) normal insulator phases. 
}
	\label{fig:growth_process}
\end{figure}

{\it Typical growth processes.}---Here, we illustrate the typical growth processes in Figs.~\ref{fig:growth_process}(a) and \ref{fig:growth_process}(b).
First, we consider the OAI phase, where the deposition at the corner is energetically favored [Fig.~\ref{fig:growth_process}(a)]. Once the corner is occupied, the adjacent site ($p=1$ in Fig.~\ref{fig:growth_process}(a)) becomes favored, followed by deposition at the site at $p=2$. Then the site at $p=3$ is also favored. The sequential depositions at $p=1,2,3$ lead to the development of a corner.
Second, we consider the normal insulator phase, where all edge sites and corner sites are equally favored for deposition [Fig.~\ref{fig:growth_process}(b)]. As an illustrative case, we consider deposition at both a corner site and another edge site. The neighboring sites, such as $q=1$ and $q=2$ in Fig.~\ref{fig:growth_process}(b), then become favored, followed by the sites on top of the previously deposited atoms ($q=3$ and $q=4$). After the sites at $q=1,2,3,4$ are occupied, the interface becomes rougher [Fig.~\ref{fig:growth_process}(b)] than that in the OAI phase [Fig.~\ref{fig:growth_process}(a)].
Thus, the OAI phase leads to the crystal shape with developed corners and with a relatively smoother interface than the normal insulator phase.
We also note that this different behavior between the normal insulator and OAI phases appears only when the Mullins-Sekerka instability is relatively strong, \textit{i.e.}, when $D_f$ is small. It is in agreement with Fig.~\ref{fig:fractal_characters}(a).

{\it Growth at high growth rates.}---In this section, we discuss the results of our simulation at a higher growth rate than the regime of $9k_{\rm B}T \leq |\Delta \mu|\leq 9.25k_{\rm B}T$. 
The OAI exhibits a dendritic morphology in the regime of large $|\Delta \mu|$ ($|\Delta \mu|>9.25k_{\rm B}T$). This behavior can be explained by the energy change associated with the deposition of a gas atom~[Eq.~\eqref{eq:energy_change_approx}]. 
The observed morphology characteristic of the OAI phase stems from the terms $\Delta {n}_{s} e_{s}+\Delta {n}_{t} e_{t}$.  
For large $|\Delta \mu|$ satisfying $|\Delta \mu|\gg \Delta {n}_{s} e_{s}+\Delta {n}_{t} e_{t}$, the growth is dominated by the term $|\Delta \mu|$.
Consequently, as $|\Delta \mu|$ increases, the morphology of the OAI increasingly resembles that of the normal insulator. 
We substantiate this observation with the values of the fractal dimension $D_f$ and the fractal dimension of coastlines $D_{f,c}$ for $|\Delta \mu|>9.25k_{\rm B}T$ in Figs.~\ref{fig:growth_shape}(c) and \ref{fig:growth_shape}(d). The value of $D_{f,c}$ in this regime is higher than that for $|\Delta \mu|\leq 9.25k_{\rm B}T$, while the value of $D_f$ for $|\Delta \mu|>9.25k_{\rm B}T$ is comparable to that in the regime of $9k_{\rm B}T \leq |\Delta \mu|\leq 9.25k_{\rm B}T$. Based on these results, we conclude that the morphology of the OAI in this high-growth-rate regime is similar to that of the normal insulator.

\begin{table}
	\centering
	\caption{Fractal dimension $D_{f}$, fractal dimension of coastlines $D_{f,c}$, and the corresponding time $t$ for Figs.~\ref{fig:fractal_characters}(c) and \ref{fig:fractal_characters}(d).
    }
	\label{tab:df_dfc_list}
     \begin{tabular}{cccc} \hline \hline
      ~~Results in Fig.~\ref{fig:fractal_characters}~~ & ~~~$D_{f}$~~~ & ~~~$D_{f,c}$~~~& ~~~$t$~~~ \\ \hline 
     (c-1) &  ~~1.78~~ & ~~1.29~~ & 1221\\
     (c-2) &  ~~1.82~~ & ~~1.41~~ & 1182\\
     (c-3) &  ~~1.80~~ & ~~1.54~~& 1123 \\
     (c-4) &  ~~1.79~~ & ~~1.22~~& 1104 \\
     (c-5) &  ~~1.75~~ & ~~1.39~~ & 1204\\
     (c-6) &  ~~1.92~~ & ~~1.50~~ & 1126\\
     (c-7) &  ~~1.69~~ & ~~1.51~~ & 1113\\ \hline 
     (d-1) &  ~~1.77~~ & ~~1.75~~ &1064 \\
     (d-2) &  ~~1.75~~ & ~~1.56~~ & 935 \\
     (d-3) &  ~~1.72~~ & ~~1.73~~ &973 \\
     (d-4) &  ~~1.75~~ & ~~1.75~~ & 925 \\
     (d-5) &  ~~1.70~~ & ~~1.90~~ & 1005 \\
     (d-6) &  ~~1.81~~ & ~~1.80~~& 939 \\
     (d-7) &  ~~1.78~~ & ~~1.73~~ & 959 \\
    \hline \hline
  \end{tabular}
\end{table}

\clearpage

\widetext
\setcounter{secnumdepth}{3}
\renewcommand{\theequation}{S\arabic{equation}}
\setcounter{equation}{0}
\renewcommand{\thetable}{S\arabic{table}}
\setcounter{table}{0}
\renewcommand{\thefigure}{S\arabic{figure}}
\setcounter{figure}{0}
\renewcommand{\thesection}{\Roman{section}}
\setcounter{section}{0}


\begin{center}
{\bf \large Supplemental Material~for 
``Hopper-Like Growth of Higher-Order Topological Insulators"}
\end{center}

\section{Filling anomaly}
We show that a two-dimensional tight-binding model  \cite{PhysRevLett.119.246402} given by Eq.~(1) in the main text exhibits the filling anomaly. 
 This model respects four-fold rotational ($C_4$) symmetry 
 \begin{align}
 C_4 \mathcal{H}(\boldsymbol{k})C_4^{-1}=\mathcal{H}(C_4 \boldsymbol{k}), \quad C_4 = \tau_{z}e^{-i\pi \sigma_z /4},
\end{align}
 and time-reversal symmetry 
 \begin{align}
 \mathcal{T} \mathcal{H}(\boldsymbol{k})\mathcal{T}^{-1}=\mathcal{H}(-\boldsymbol{k}), \quad \mathcal{T}=i\sigma_y K,
 \end{align}
with $K$ being the complex conjugate and $\mathcal{T}^2=-1$, and therefore the Hamiltonian $\mathcal{H}(\boldsymbol{k})$ belongs to class AII~\cite{PhysRevB.55.1142}. 
The filling anomaly of a two-dimensional system indicates the mismatch between the total charge of electrons and ions in finite systems with rotational symmetry, and it leads to corner charges \cite{PhysRevB.99.245151}. 
When the positions of the ions are at the center of the unit cell, the formula of the corner charge in $C_4$-symmetric crystals in class AII is given by \cite{PhysRevB.103.205123}
\begin{align}
	Q_{c }^{(4)} \equiv \frac{|e|}{4}(n^{(\rm ion)}-\nu)+\frac{|e|}{2}[M_{1}^{(4)}]\ \ ({\rm mod}\ 2e),
\end{align}
where $n^{\rm (ion)}|e|$ is an ionic charge at the center of the unit cell, $\nu$ is the number of occupied bands, and $[M_{1}^{(4)}]$ is the difference of the numbers of occupied states with the $C_4$ eigenvalue $e^{i\frac{\pi}{4}}$ between the $M$ point 
($\boldsymbol{k}=(\pi,\pi)$) and the $\Gamma$ point ($\boldsymbol{k}=(0,0)$) in the $\boldsymbol{k}$ space.
Although perturbations preserving $C_4$ symmetry can change the corner charge by integer multiples of $e$ \cite{PhysRevB.102.165120}, we focus only on changes of the corner charge caused by perturbations on electronic states in this work. Consequently, we define the filling anomaly modulo 2$e$ due to the Kramers theorem \cite{PhysRevResearch.1.033074, kooi2021bulk}. 

In the Hamiltonian $\mathcal{H}(\boldsymbol{k})$ with the parameters satisfying $|m|<2t$ ($t>0$), the numbers of occupied states with the $C_4$ eigenvalue $e^{i\frac{\pi}{4}}$ at the $M$ and $\Gamma$ points are given by
\begin{align}
	n_{\frac{\pi}{4}}(M) = 2, \quad n_{\frac{\pi}{4}}(\Gamma) = 0,
\end{align} 
respectively.
Therefore, the corner charge is given by 
\begin{align}
	Q_{c }^{(4)} \equiv |e|\ \ ({\rm mod}\ 2e),
\end{align}
which indicates that our model exhibits the nonzero filling anomaly. This value does not change as long as $|m|<2t$ ($t>0$) is satisfied, and the quantized corner charge is topologically protected. 

\section{Transformation to the doubled BBH model}
In this section, we show that the simple two-dimensional tight-binding model $\mathcal{H}(\boldsymbol{k})$ given by Eq.~(1) in the main text is adiabatically connected to the direct sum of two Benalcazar-Bernevig-Hughes (BBH) models \cite{benalcazar2017quantized}, $\mathcal{H}_{\rm BBH}(\boldsymbol{k}) \oplus  \mathcal{H}_{\rm BBH}(\boldsymbol{k})$ by local transformations.
To begin with, by using a unitary matrix $U_{\mu}=(\mu_y + \mu_z)/\sqrt{2}$, we transform the original model $\mathcal{H}(\boldsymbol{k})$ as 
\begin{align}
\mathcal{H}(\boldsymbol{k})\  \rightarrow \  & U_{\mu}^{\dagger} \mathcal{H}(\boldsymbol{k})U_{\mu} 
 =  \begin{pmatrix}
		\mathcal{H}_{\tau \sigma+}(\boldsymbol{k}) & \\
		& \mathcal{H}_{\tau \sigma-}(\boldsymbol{k})
	\end{pmatrix},
\end{align}
where the diagonal parts are defined as 
\begin{align}
	\mathcal{H}_{\tau \sigma \pm}(\boldsymbol{k}) :=& \biggl( m-t\sum_{i=x,y}\cos k_i\biggr) \tau_z  +v   \tau_x  \biggl( \sum_{i=x,y} \sin k_i \sigma_i \biggr)  \pm \Delta(\cos k_x - \cos k_y)\tau_y \sigma_0.
\end{align}
Thanks to this transformation, we can decompose the $8\times 8$ Hamiltonian $\mathcal{H}(\boldsymbol{k})$ to the two $4 \times 4$  Hamiltonians $\mathcal{H}_{\tau \sigma +}(\boldsymbol{k})$ and $\mathcal{H}_{\tau \sigma -}(\boldsymbol{k})$. 
In the following,  we show that $\mathcal{H}_{\tau \sigma \pm}(\boldsymbol{k})$ with specific values of parameters can be connected to the BBH model via unitary transformations. For our purpose, we introduce the following unitary matrix 
\begin{align}
		U_{\tau \sigma} = \frac{1}{2}\begin{pmatrix}
			1 & -i & 1 & -i \\
			i & -1 & i & -1 \\
			1 & -i & -1 & i \\
			-i & 1 & i & -1 
		\end{pmatrix},
\end{align}
which acts on the $\tau_i \sigma_j$ subspace.  Here, the tensor product is defined by $AB=(A_{ij}B)_{ij}$.
Under the unitary transformation $\hat{U}_{\tau \sigma}$, the following matrices transform as
\begin{align}
	&\tau_z\sigma_0  \rightarrow U^\dagger_{\tau \sigma} \tau_z\sigma_0 U_{\tau \sigma}=\tau_x \sigma_0, \\
	&\tau_x \sigma_x  \rightarrow  U^\dagger_{\tau \sigma} \tau_x \sigma_x U_{\tau \sigma}=   -\tau_y\sigma_z, \\
	&\tau_x \sigma_y \rightarrow U^\dagger_{\tau \sigma} \tau_x \sigma_y U_{\tau \sigma}= -\tau_y \sigma_x, \\
	&\tau_y \sigma_0 \rightarrow U^\dagger_{\tau \sigma} \tau_y \sigma_0 U_{\tau  \sigma}= -\tau_y \sigma_y,
\end{align}
 and therefore this unitary transformation maps $\mathcal{H}_{\tau \sigma \pm}(\boldsymbol{k})$ to
\begin{align}
	 \mathcal{H}'_{\tau \sigma \pm}(\boldsymbol{k}) = & U^\dagger_{\tau \sigma} \mathcal{H}_{\tau \sigma \pm}(\boldsymbol{k})  U_{\tau \sigma} \nonumber \\
	 =	  & \biggl( m-t\sum_{i=x,y}\cos k_i\biggr) \tau_x   -v \sin k_x  \tau_y \sigma_z  -v \sin k_y  \tau_y \sigma_x \mp \Delta(\cos k_x - \cos k_y)\tau_y \sigma_y.
\end{align}
Our goal is to transform $\mathcal{H}_{\tau \sigma \pm}'(\boldsymbol{k})$ into the BBH model \cite{benalcazar2017quantized}
\begin{align}
	\mathcal{H}_{\rm BBH}(\boldsymbol{k})= (\gamma+\lambda \cos k_x) \Gamma_4 +\lambda \sin k_x \Gamma_3 +(\gamma + \lambda \cos k_y)\Gamma_2 +\lambda \sin k_y \Gamma_1, 
\end{align}
where  $\{ \Gamma_i \}$ are the sets of Dirac matrices given by  $\Gamma_1= -\tau_y \sigma_x$,  $\Gamma_2 = -\tau_y \sigma_y$, $\Gamma_3 = -\tau_y \sigma_z$, and $\Gamma_4 = \tau_x \sigma_0$.
To relate $\mathcal{H}'_{\tau \sigma +}(\boldsymbol{k})$ to $\mathcal{H}_{\rm BBH}(\boldsymbol{k})$, we choose the parameters as follows:
\begin{gather}
	\sqrt{2} t = v = \sqrt{2} \Delta = \lambda , \quad m=\gamma. \label{eq:parameter_to_BBH}
\end{gather}
With these parameters and the sets of Dirac matrices $\{ \Gamma_i \}$, the Hamiltonian $\mathcal{H}'_{\tau \sigma \pm}(\boldsymbol{k})$ can be written as 
\begin{align}
	\mathcal{H}'_{\tau \sigma \pm}(\boldsymbol{k}) &= \biggl(\gamma-\frac{\lambda}{\sqrt{2}}(\cos k_x + \cos k_y )\biggr) \Gamma_4 +\lambda \sin k_x \Gamma_3  \nonumber \\
&+\lambda \sin k_y  \Gamma_1  \pm \frac{\lambda}{\sqrt{2}}(\cos k_x - \cos k_y)\Gamma_2.
\end{align}
Furthermore, we consider the limit $\gamma/ \lambda \rightarrow 0$, where the BBH model is topologically nontrivial.
In this limit, the models $\mathcal{H}'_{\tau \sigma \pm}(\boldsymbol{k})$ and $\mathcal{H}_{\rm BBH}(\boldsymbol{k})$ reduce to 
\begin{align}
	\mathcal{H}'_{\tau \sigma \pm}(\boldsymbol{k}) = &  \lambda \cos k_x \biggl( \pm \frac{\Gamma_2}{\sqrt{2}} -\frac{\Gamma_4}{\sqrt{2}} \biggr) + \lambda \cos k_y \biggl( \mp \frac{\Gamma_2}{\sqrt{2}} -\frac{\Gamma_4}{\sqrt{2}} \biggr)+\lambda \sin k_x \Gamma_3 +\lambda \sin k_y  \Gamma_1, 
\end{align}
and
\begin{align}
		 \mathcal{H}_{\rm BBH}(\boldsymbol{k})= & \lambda \cos k_x \Gamma_4 + \lambda \cos k_y \Gamma_2 +\lambda \sin k_x \Gamma_3  +\lambda \sin k_y \Gamma_1,
\end{align}
respectively. We introduce the rotation matrix in the $\Gamma_{2,4}$ subspace
\begin{align}
	R(\theta)= \exp \Bigl( \frac{i}{2} \theta \Sigma_{24} \Bigr),
\end{align}
with
$
	 \Sigma_{24} =\frac{i}{4}[\Gamma_2, \Gamma_4]
$.
Under the transformation with $\theta = -3\pi/2$,  $\Gamma_2$ and $\Gamma_4$  transform as 
\begin{align}
		\begin{pmatrix}
			\Gamma_2 \\
			\Gamma_4
		\end{pmatrix}
		\rightarrow
		& \begin{pmatrix}
			R^\dagger(-3\pi/2)\Gamma_2 R(-3\pi/2) \\
			R^\dagger(-3\pi/2)\Gamma_4 R(-3\pi/2)
		\end{pmatrix}
		\nonumber \\
		 &=
			\begin{pmatrix}
			-\frac{1}{\sqrt{2}} &   \frac{1}{\sqrt{2}} \\
			-\frac{1}{\sqrt{2}} &  - \frac{1}{\sqrt{2}}
		\end{pmatrix}
		\begin{pmatrix}
			\Gamma_2 \\
			\Gamma_4
		\end{pmatrix}.
\end{align}
With this rotation matrix $R(-3\pi/2)$ and the parameters in Eq.~(\ref{eq:parameter_to_BBH}) in the limit $\gamma/\lambda \rightarrow 0$, it follows that
\begin{align}
	R^\dagger \Bigl( -\frac{3\pi}{2} \Bigr) \mathcal{H}'_{\tau \sigma+}(\boldsymbol{k}) R \Bigl( -\frac{3\pi}{2} \Bigr)=\mathcal{H}_{\rm BBH}(\boldsymbol{k}).
\end{align}
Also, the Hamiltonian $\mathcal{H}_{\tau \sigma-}'(\boldsymbol{k})$ satisfies the following equation:
\begin{align}
	 \biggl[R\Bigl( -\frac{3\pi}{2} \Bigr) \frac{\Gamma_2 + \Gamma_4}{\sqrt{2}}  \biggr]^\dagger \mathcal{H}_{\tau \sigma-}'(\boldsymbol{k}) \biggl[R\Bigl( -\frac{3\pi}{2} \Bigr) \frac{\Gamma_2 + \Gamma_4}{\sqrt{2}}  \biggr] 
	=\mathcal{H}_{\rm BBH}(\boldsymbol{k}).
\end{align}
To summarize the above calculations, with the parameters in Eq.~(\ref{eq:parameter_to_BBH}) and in the limit $\gamma/\lambda \rightarrow 0$,  the original Hamiltonian $\mathcal{H}(\boldsymbol{k})$ satisfies
\begin{align}
U_{\rm BBH}^{\dagger}
	 \mathcal{H}(\boldsymbol{k})
U_{\rm BBH}
	  =& \begin{pmatrix}
		\mathcal{H}_{\rm BBH}(\boldsymbol{k}) & \\
		& \mathcal{H}_{\rm BBH}(\boldsymbol{k}) 
	\end{pmatrix},
\end{align}
where a unitary matrix $U_{\rm BBH}$ is defined as
\begin{align}
U_{\rm BBH} :=  U_\mu
 &\begin{pmatrix}
		{U}_{\tau \sigma} R\bigl(-\frac{3\pi}{2} \bigr)   & \\
		& {U}_{\tau \sigma} R\Bigl(-\frac{3\pi}{2} \Bigr)  \frac{\Gamma_2 + \Gamma_4}{\sqrt{2}} 
	\end{pmatrix} .
\end{align}
The BBH model is topologically nontrivial in $|\gamma/\lambda|<1$ and trivial in $|\gamma/\lambda|>1$, and 
the topological phase transition occurs at  $|\gamma/\lambda|=1$ \cite{benalcazar2017quantized}.
Thus, the Hamiltonian $\mathcal{H}(\boldsymbol{k})$ with the parameter values in Eq.~(\ref{eq:parameter_to_BBH}) satisfying $|\gamma/\lambda|<1$ can be adiabatically connected to the doubled BBH model $\mathcal{H}_{\rm BBH}(\boldsymbol{k}) \oplus \mathcal{H}_{\rm BBH}(\boldsymbol{k})$ in the OAI phase, and then $\mathcal{H}(\boldsymbol{k})$ is topologically equivalent to the doubled BBH model.

\section{Details of calculation of $\Delta n_s$ and $\Delta n_t$}
We give formulae to determine $\Delta n_s$ and $\Delta n_t$ from the geometry of a crystal in 2D real space.
\subsection{Review of Wannier function}
To begin with, we review the Wannier functions~\cite{RevModPhys.84.1419}, which capture the number of in-gap states in the following.
The Wannier functions for multiple bands with degeneracy are defined as 
\begin{equation}
	\ket{w_{\alpha \boldsymbol{R}}} := \frac{1}{\sqrt{N_u}}\sum_{\boldsymbol{k}} e^{-i\boldsymbol{k}\cdot \boldsymbol{R}} \ket{\tilde{\psi}_{\alpha \boldsymbol{k}}},
\end{equation}
where $\alpha$ is the band index, $\boldsymbol{R}$ is a lattice vector, $N_{u}$ is the number of unit cells, and $\ket{\tilde{\psi}_{\alpha \boldsymbol{k}}}$ is a Bloch-like function that is smooth everywhere in the Brillouin zone. Here, the wavefunction $\ket{\tilde{\psi}_{\alpha \boldsymbol{k}}}$ for an isolated group of $J$ bands that is not degenerate with other bands is given by 
\begin{align}
	\ket{\tilde{\psi}_{\alpha \boldsymbol{k}}} = \sum^{J}_{\beta=1}U_{\beta \alpha}(\boldsymbol{k})\ket{{\psi}_{\beta \boldsymbol{k}}},
\end{align}
where $\ket{{\psi}_{\beta \boldsymbol{k}}}$ is the Bloch wavefunction and $U(\boldsymbol{k})$ is a unitary matrix of dimension $J$ that is periodic in the Brillouin zone.
The gauge of the wave functions depends on the unitary matrix $U(\boldsymbol{k})$. The unitary matrix is chosen in such a way that the transformed states $\ket{\tilde{\psi}_{\alpha \boldsymbol{k}}}$ are smooth everywhere in the Brillouin zone. This choice leads to a well-localized Wannier function around $\boldsymbol{R}$. 

We introduce Wannier centers, defined as the expectation values of the position $\hat{\boldsymbol{r}}$ in the unit cell, which can be regarded as centers of the Wannier functions. The Wannier centers are given by 
\begin{equation}
    \bar{\boldsymbol{r}}_{\alpha}=\bra{w_{\alpha \boldsymbol{R}}}\hat{\boldsymbol{r}}\ket{w_{\alpha \boldsymbol{R}}}=\int \frac{d^2\boldsymbol{k}}{4\pi^2} \boldsymbol{\mathcal{A}}_{\alpha}(\boldsymbol{k})+\boldsymbol{R},
\end{equation}
where $\boldsymbol{\mathcal{A}}_{\alpha}$ is the multi-band Berry connection for $J$ bands defined as
\begin{align}
	\boldsymbol{\mathcal{A}}_{\alpha}(\boldsymbol{k}):=i\bra{\tilde{u}_{\alpha \boldsymbol{k}}} \partial_{\boldsymbol{k}} 
	\ket{\tilde{u}_{\alpha \boldsymbol{k}}},
\end{align} 
where $ \ket{\tilde{u}_{\alpha \boldsymbol{k}}}$ is the periodic part of the Bloch-like function $\ket{\tilde{\psi}_{\alpha \boldsymbol{k}}}$.

\subsection{Positions of Wannier centers}

\begin{figure}
	\includegraphics[width=0.5\columnwidth]{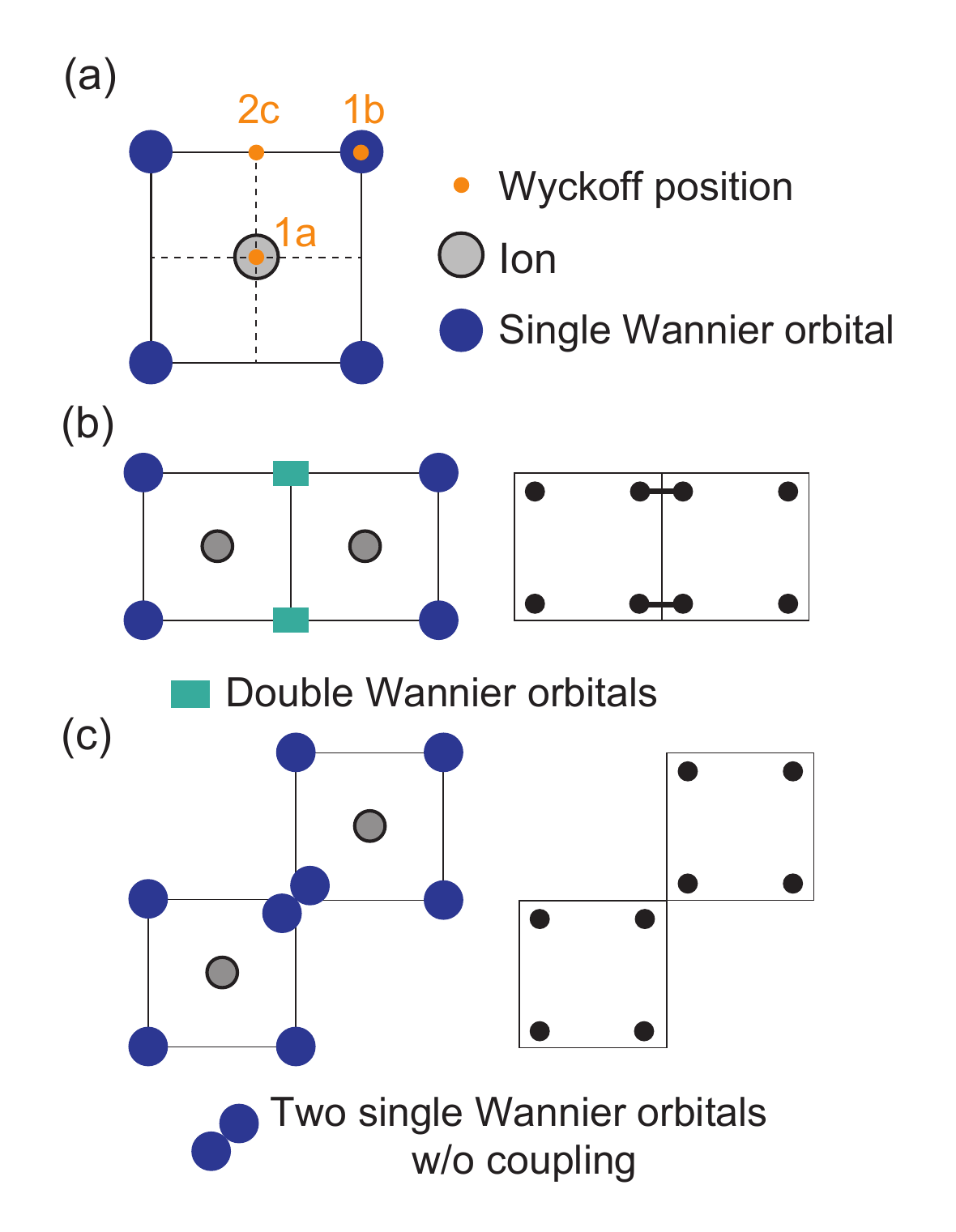}
	\caption{(a) Positions of an atom and Wannier centers and Wyckoff positions in the unit cell. (b) Double Wannier orbitals between the two unit cells. The overlapping Wannier orbitals correspond to the bonding of Wannier orbitals between the two adjacent atoms. (c) Two single Wannier orbitals without coupling. The Wannier orbitals between next-nearest-neighbor atoms do not form a bond.}
	\label{fig:wyckoff_position}
\end{figure} 

Next, we discuss the positions of the Wannier centers in terms of crystal symmetries.  
In the presence of crystal and time-reversal symmetries, when the set of  Wannier functions forms a representation of the symmetry group, these Wannier functions are called ``symmetric.'' Here, we assume that the Wannier functions are symmetric, and therefore the Wannier centers are also symmetric. Then, the positions of Wannier centers are classified by the Wyckoff positions, which are points in the unit cell classified in terms of the site symmetry groups.
The site symmetry group for a given point $\boldsymbol{r}$ is defined as the group that consists of all symmetry operations leaving $\boldsymbol{r}$ invariant. Here, we assume the wallpaper group $p4$ generated by a four-fold rotation, because our model $\mathcal{H}(\boldsymbol{k})$ in the main text respects four-fold rotational symmetry.
The maximal Wyckoff positions for the wallpaper group $p4$ are given by 1a, 1b, and 2c
[Fig.~\ref{fig:wyckoff_position}(a)]. The positions 1a and 1b are the center and corner of the unit cell, respectively. 
In the model $\mathcal{H}(\boldsymbol{k})$ in the main text, the ions are located at the position 1a, and the four Wannier orbitals are localized at the position 1b \cite{PhysRevLett.119.246402}. 

\begin{figure*}
	\includegraphics[width=0.85\columnwidth]{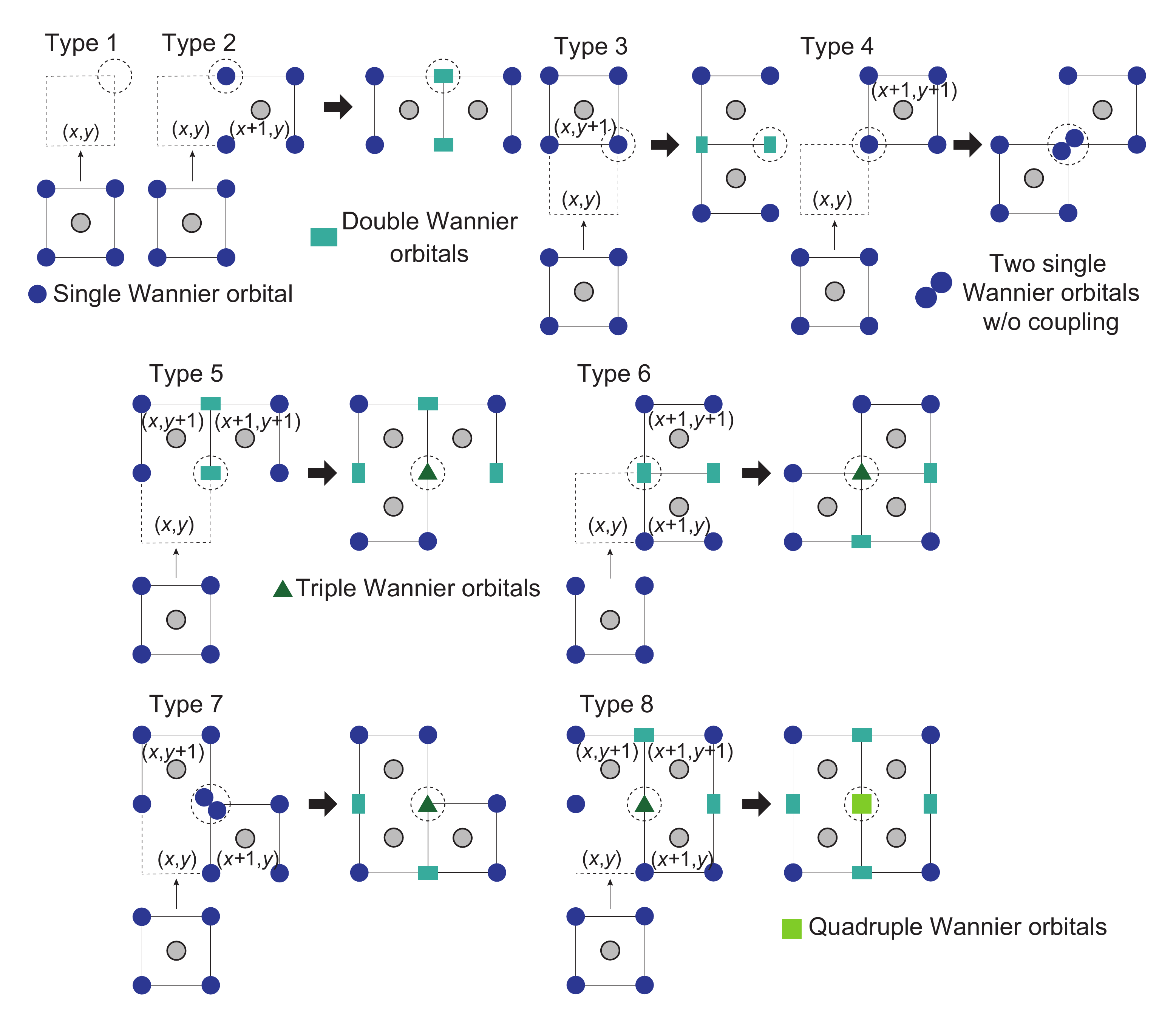}
	\caption{Deposition of an atom at $(x,y)$. The eight types are classified in terms of the presence and absence of atoms at $(x+1,y)$, $(x, y+1)$, and $(x+1,y+1)$. The dotted circles indicate the positions of the corners at $(x+1/2,y+1/2)$.}
	\label{fig:solidification_pattern}
\end{figure*} 

We consider a 2D covalent crystal with covalent bonds formed by pairs of Wannier functions. 
We regard the overlapping Wannier orbitals from neighboring atoms to form bonding orbitals, and then the system is a covalent crystal where the orbitals are located near the four corners of each unit cell [Fig.~\ref{fig:wyckoff_position}(b)]. 
Here, we assume that the covalent bonds are formed only between neighboring atoms displaced along the $x$ and $y$ directions because $\mathcal{H}(\boldsymbol{k})$ has only the nearest neighbor hoppings. Thus, the Wannier orbitals from two neighboring atoms mutually displaced along $[11]$ or $[1\bar{1}]$ directions do not form a bond [Fig.~\ref{fig:wyckoff_position}(c)], resulting in the two single Wannier orbitals rather than one double Wannier orbital. 
As we discuss in the main text, the double and quadruple Wannier orbitals result in stable states, and the single Wannier orbital and the triple Wannier orbitals lead to unstable states in the band gap.
Thus, the single Wannier orbitals and triple Wannier orbitals contribute to the number of in-gap states. 

\subsection{Numbers of Wannier centers}
We explain how the deposition of atoms changes the number of single Wannier orbitals and triple Wannier orbitals. 
We consider the upper right corner $(x+1/2,y+1/2)$ of the unit cell at $(x,y)$, and we introduce numbers, $n_{s}(x+1/2,y+1/2)$ and $n_{t}(x+1/2,y+1/2)$, which are the number of single Wannier orbitals and that of triple Wannier orbitals at $(x+1/2,y+1/2)$,  respectively. 
We consider the changes in $n_{s}(x+1/2,y+1/2)$ and $n_{t}(x+1/2,y+1/2)$ during the deposition of an atom at $\boldsymbol{R}=(x,y)$.
We classify the depositions into eight types shown in Fig.~\ref{fig:solidification_pattern}, depending on the presence and absence of atoms at the neighboring sites before the deposition of an atom at $(x,y)$ [dotted circles in Fig.~\ref{fig:solidification_pattern}]. 
Then, the changes in $n_{s}(x+1/2,y+1/2)$ and $n_{t}(x+1/2,y+1/2)$ upon the deposition,  $\Delta n_{s}(x+1/2,y+1/2)$ and $\Delta n_{t}(x+1/2,y+1/2)$, are summarized in Table~\ref{table:delta_nm}.

The above results indicate that $\Delta n_{s}(x,y)$ and $\Delta n_{t}(x,y)$  are determined only by the information of the presence and absence of atoms in the neighboring unit cells. 
The changes in the number of single and triple Wannier orbitals during the deposition of an atom at ($x,y$) are respectively given by 
\begin{align}
	&\Delta n_{s}(x,y) =\sum_{\alpha, \beta=\pm1} \Delta n_{s}(x+\tfrac{\alpha}{2},y+\tfrac{\beta}{2}),\\
	&\Delta n_{t}(x,y) =\sum_{\alpha, \beta=\pm1} \Delta n_{t}(x+\tfrac{\alpha}{2},y+\tfrac{\beta}{2}).
\end{align}
Both $\Delta n_s (x,y)$ and $\Delta n_t (x,y)$ contribute to the in-gap states. 
By using these equations and the results in Table~\ref{table:delta_nm}, we can obtain the third and fourth terms in the approximation $\Delta E_{n} \simeq  \Delta n_b e_b + \mu_s +\Delta {n}_{s} e_{s}+\Delta {n}_{t} e_{t}$ in the main text. 

\begin{table*}
\begin{tabular}{c c c c c c}
\hline
\hline
Type & 
$n(x+1,y)$\ \ \   &  $n(x,y+1)$\ \ \  & $n(x+1,y+1)$\ \ \  & $\Delta n_{s} (x+1/2, y+1/2)$\ \ \   &  $\Delta n_{t} (x+1/2, y+1/2)$ \\ 
\hline
Type 1 & 0& 0 & 0 & 1 & 0 
\\
Type 2 & 1& 0 & 0 & $-1$ & 0 
\\
Type 3 & 0& 1 & 0 & $-1$ & 0 
\\
Type 4 & 0& 0 & 1 & $1$ & 0 
\\
Type 5 & 0& 1 & 1 & 0 & 1 
\\
Type 6 & 1& 0 & 1 & 0 & 1 
\\
Type 7 & 1& 1 & 0 & $-2$ & 1 
\\
Type 8 & 1& 1 & 1 & 0 & $-1$ 
\\
\hline
\hline
\end{tabular}
\caption{The changes in the numbers of single Wannier orbitals and triple Wannier orbitals at $(x+1/2,y+1/2)$, $\Delta n_{s} (x+1/2,y+1/2)$ and $\Delta n_{t} (x+1/2,y+1/2)$, via deposition of one atom onto the site at $(x,y)$. 
Here, $n(x,y)$ is the number of atoms at  $(x,y)$ and takes $0$ or $1$.
The types in the first column correspond to the deposition types shown in Fig.~\ref{fig:solidification_pattern}.  }
\label{table:delta_nm}
\end{table*}

\section{Monte Carlo simulation of the crystal growth}
In this section, we show the details of our Monte Carlo simulation of crystal growth. 
We consider the first quadrant ($x \geq 0$ and $y \geq 0$) of the square lattice $\boldsymbol{R}=(x,y)$ and $x, y \in \mathbb{Z}$, where we impose the periodic boundary condition connecting the positive $x$-axis and the positive $y$-axis. 
We assume the vessel for the gas has a circular shape with its center at the origin $(0,0)$, and its volume is denoted by $\Omega$ measured in the unit of the unit cell of the solid crystal.  
Each lattice site at $\boldsymbol{R}=(x,y)$ can exhibit one of three states: occupied by a gas atom, occupied by a solid atom, or vacant. 
Let $N_g$ and $N_s$ denote the numbers of gas atoms and solid atoms, respectively, and let $N_{g}^0$ and $N_s^0$ be their initial values. 
At the beginning of the growth, a square solid nucleus (seed crystal) consisting of $N_{s}^0 := L_n \times L_n$ atoms is located at the region with $0 \leq x \leq L_n-1$ and $0 \leq y \leq L_n -1$, and the size of the seed crystal is set to $L_n=20$. 
Meanwhile, $N_g$ gas atoms are randomly arranged so that the initial gas density is given by $n_g  =N^0_g/V_g= 0.5$, where $V_g = \Omega - N_s^0$ is the initial volume of the vessel outside of the solid nucleus. In the simulation, the gas atoms move freely in the vessel to contribute to the entropy
\begin{align}
	S_{g} =k_{\rm B} \ln \left[ \frac{(\Omega- N_{s})!}{(N_{t}-N_{s})! (\Omega - N_{t})!} \right],
\end{align}
where $N_t=N_g+N_s$ is the total number of atoms in the vessel.
Furthermore, we get the total free energy
\begin{align}
	F(N_{s}) = E_{s}(N_{s}) + \mu_{s} N_{s}  +\mu_{g,c} (N_{t}-N_s) -T S_{g},
\end{align}
where $E_{s}(N_{s})$ is the energy of the solid measured from the chemical potential, $\mu_s$ is the chemical potential of the solid phase, and $\mu_{g,c}$ $(<0)$ denotes the energy per gas atom. 

With Stirling's approximation, the partial derivative of the total free energy with respect to $N_s$ is given by 
\begin{align}
    \frac{\partial F(N_{s})}{\partial N_s} \approx  \frac{\partial E(N_{s})}{\partial N_{s}} + \mu_{s} -\mu_g, 
\end{align}
where $\mu_g$ is the chemical potential of the gas phase and given by 
\begin{align}\label{eq:def_mug}
	\mu_g = \mu_{g,c} +k_{\rm B} T \ln n_{g}.
\end{align}

The process of the Monte Carlo simulation of crystal growth consists of the following steps.  
To begin with, we randomly select a single atom among all the gas atoms and the interfacial solid atoms. 
Here, we define a solid atom as interfacial if at least one of its adjacent sites is not occupied by a solid atom. 
If the chosen atom is a gas atom, it moves randomly to an unoccupied nearest-neighbor site. Subsequently, if the gas atom comes into contact with a solid atom after this movement, it is deposited on the surface with a probability $W$.
On the other hand, if the chosen atom is an interfacial solid atom, evaporation of the solid atom occurs with a probability $W'$, where the atom turns into a gas atom. 
Here, we assume that surface diffusion is negligible compared to the growth rate, and therefore surface diffusion is not considered in our model.

Because the process of deposition and evaporation at the interface should satisfy the detailed balance condition, we use the heat-bath algorithm: The transition probabilities $W$ and $W'$ are defined as 
\begin{gather}
	W:=\frac{1}{1+e^{\Delta E/k_{\rm B}T}}, \\
	W':=\frac{1}{1+e^{-\Delta E/k_{\rm B}T}},
\end{gather}
 where $\Delta E$ is the energy change upon the deposition of an atom and is given by
 \begin{align}
     \Delta E= \Delta E_{n} -\mu_g & \simeq \Delta n_b e_b  +\Delta {n}_{s} e_{s}+\Delta {n}_{t} e_{t}+ \mu_s-\mu_g \nonumber \\
     &=\Delta n_b e_b +\Delta {n}_{s} e_{s}+\Delta {n}_{t} e_{t} +\Delta \mu - k_{\rm B}T \ln n_g.
 \end{align}
 Here, we approximate $\Delta E_{n}$ by Eq.~(\ref{eq:energy_change_approx}), which enables us to perform the simulation without calculating the Hamiltonian $\mathcal{H}(\boldsymbol{k})$ in each step. 

To see a steady growth of the crystal, an open system is preferred where $\Omega$ and $N_{t}$ are infinite. To emulate an open system, we position a reservoir of gas atoms at the periphery of the system with a thickness $d_r = 5$, where the gas density is maintained at the gas density $N_g^r/V_g^r = 0.5$ with $N_g^r$ being the number of gas atoms in the reservoir and $V_g^r$ being the volume of the reservoir.
Furthermore, for the steady growth of the crystal, if the density of the gas in the region of thickness $d_c=5$ in front of the reservoir decreases by 3$\%$, we deem the crystal to be too close to the reservoir and subsequently move the reservoir outward.  
Additionally, we introduce a time increment $\Delta t = (4N_g)^{-1}$ for each movement of a gas atom in order to measure time in the simulation. This time increment implies that each gas atom moves once on average during $N_g \Delta t =1/4$, and consequently, the diffusion constant is set to $D=1$. 
Consequently, each interfacial solid atom tries to evaporate once on average in a time increment of $N_g \Delta t =1/4$.

\section{$e^{(\mathrm{n})}_b=e_b^{(\mathrm{OAI})}$ case}

\begin{figure}[b]
    \includegraphics[width=1.\columnwidth]{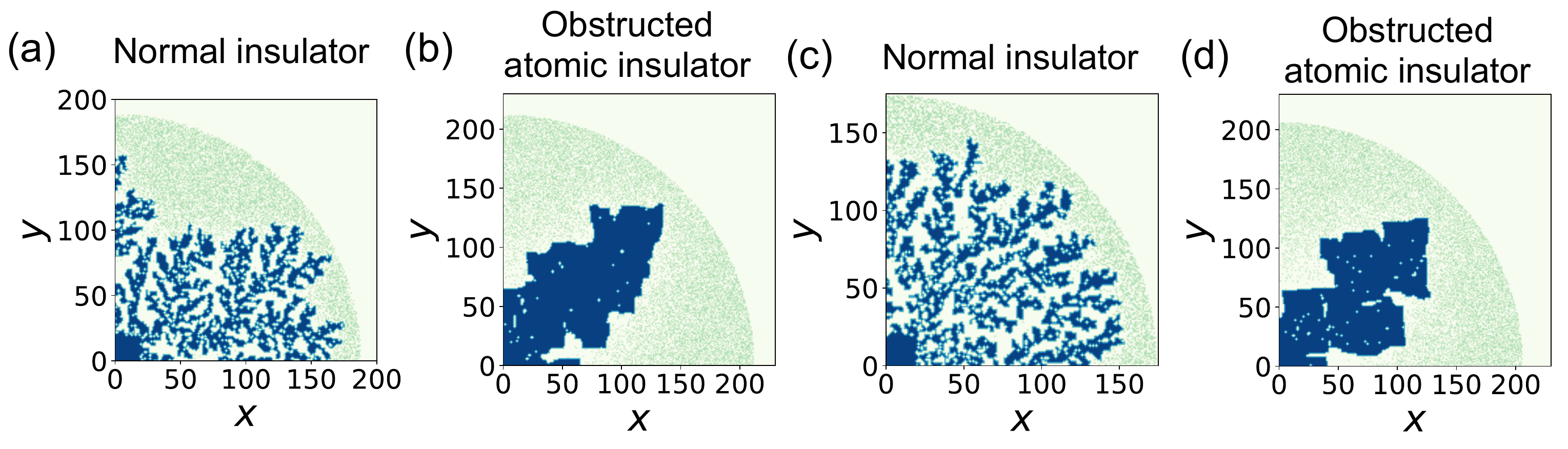}
	\caption{(a)-(d)~Crystal shapes obtained from the simulations for (a,b) $\Delta \mu/k_{\rm B}T = -9$ and (c,d) $\Delta \mu/k_{\rm B}T = -9.25$. 
    The blue, light blue, and light green sites indicate the solid atoms, interfacial solid atoms, and gas atoms, respectively. 
    The parameters are set to (i)~$e_b=0.6439$, $e_s = 0$, $e_t = 0$ for the normal insulator~[(a) and (c)] and (ii)~$e_b=0.6439$, $e_s = 0.2413$, $e_t = 0.1488$ for the obstructed atomic insulator~[(b) and (d)]. The corresponding times are (a)~$t=173$, (b)~$t=159$, (c)~$t=1500$, and (d)~$t=1400$.
    We choose $k_{\rm B}T=0.16$ in both the (i) and (ii) cases.}
	\label{fig:eb_choice_SM}
\end{figure}

In this section, we perform additional simulations with $e_b$ set to the same value for both the normal insulator and the obstructed atomic insulator.
We perform the simulation with two types of parameters:~(i) the normal insulator case:
\begin{align}
    (e_b,e_s,e_t)=(e^{(\mathrm{n})}_b, 0, 0),    
\end{align}
and (ii) the obstructed atomic insulator case:
\begin{align}
    (e_b,e_s,e_t)=(e_b^{(\mathrm{OAI})}, e_s^{(\mathrm{OAI})}, e_t^{(\mathrm{OAI})}).
\end{align} 
In the main text, we choose the value of $e^{(\mathrm{n})}_b$ to be $e^{(\mathrm{n})}_b=e_b^{(\mathrm{OAI})} + e_s^{(\mathrm{OAI})} + e_t^{(\mathrm{OAI})}$. 
In this section, we alternatively put the value of $e^{(\mathrm{n})}_b$ to be
\begin{align}
    e^{(\mathrm{n})}_b=e_b^{(\mathrm{OAI})}.
\end{align}
Figure~\ref{fig:eb_choice_SM} shows the crystal morphologies obtained for (i)~$e_b=0.6439$, $e_s = 0$, $e_t = 0$ (the normal insulator)~[Figs.~\ref{fig:eb_choice_SM}(a) and \ref{fig:eb_choice_SM}(c)] and (ii)~$e_b=0.6439$, $e_s = 0.2413$, $e_t = 0.1488$ (the obstructed atomic insulator)~[Figs.~\ref{fig:eb_choice_SM}(b) and \ref{fig:eb_choice_SM}(d)]. 
These results demonstrate that the crystal morphologies of the normal insulator differ significantly from those of the obstructed atomic insulator. 
Specifically, the normal insulator exhibits many irregular branches in contrast to the relatively smooth interface of the obstructed atomic insulator.

\begin{figure}
    \includegraphics[width=1.\columnwidth]{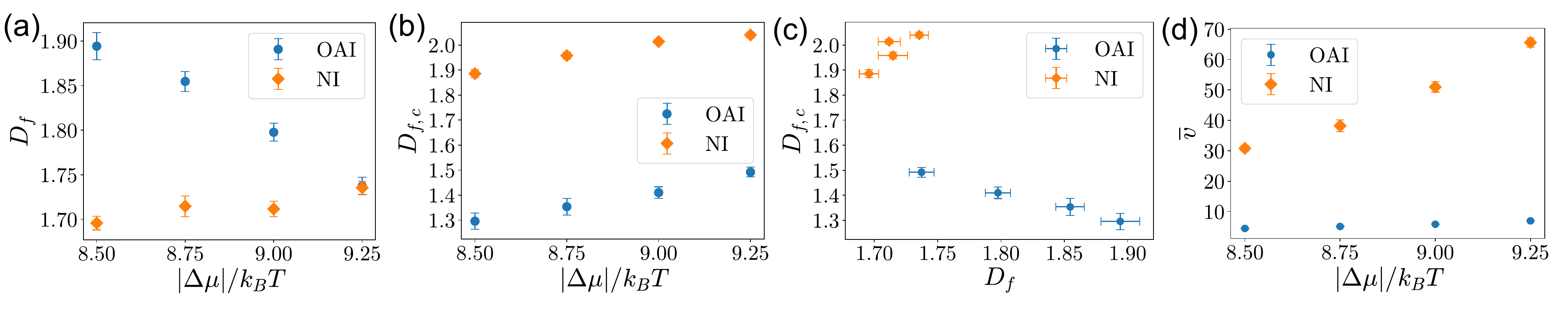}
	\caption{(a)~Fractal dimensions $D_f$. (b)~Fractal dimension of coastlines $D_{f,c}$. (c)~Fractal dimensions $D_f$ versus fractal dimensions of coastlines $D_{f,c}$. (d)~Growth rate $\bar{v}$ versus $|\Delta \mu|/k_{\rm B}T$. Here, ``NI'' and ``OAI'' indicate the parameters $e_b=0.6439$, $e_s = 0$, $e_t = 0$ (i) for the normal insulator and (ii) $e_b=0.6439$, $e_s = 0.2413$, $e_t = 0.1488$ for the obstructed atomic insulator, respectively.
    The results for the obstructed atomic insulator are reproduced from those in the main text. 
    For the normal insulator, the error bars indicate the standard error of the mean of ten simulations for each value of $|\Delta \mu|/k_{\rm B}T$. The fits for the fractal dimensions are performed only for data points with $N_{s,i}>100$.
    }
	\label{fig:eb_choice2_SM}
\end{figure}

The morphological difference between these two phases is quantified by both the fractal dimension $D_f$ and the fractal dimension of coastlines $D_{f,c}$.
Figure~\ref{fig:eb_choice2_SM}(a) shows that for $|\Delta \mu|/k_{\rm B}T< 9.25$, the fractal dimension $D_f$ of the normal insulator is smaller than that of the obstructed atomic insulator. The fractal dimension of coastlines $D_{f,c}$ of the normal insulator is higher than that of the obstructed atomic insulator~[Fig.~\ref{fig:eb_choice2_SM}(b)].
Consequently, the normal insulator exhibits a fractal morphology characterized by $D_f<1.75$ featuring many irregular branches ($D_{f,c}>1.8$), in contrast to the obstructed atomic insulator~[Fig.~\ref{fig:eb_choice2_SM}(c)].
This morphology, characterized by irregular branches, exhibits a higher value of $D_{f,c}$ than that of Region 2 in Fig.~3(a) of the main text and corresponds to a diffusion-limited-aggregation (DLA)-like pattern~\cite{Witten1981, Witten1983}. Consequently, it does not correspond to any of the three regions in Fig.~3(a).
This morphological difference stems from the significantly higher growth rate $\bar{v}$ of the normal insulator compared to the obstructed atomic insulator~[Fig.~\ref{fig:eb_choice2_SM}(d)]. As discussed in the main text, sufficiently high growth rates trigger the Mullins-Sekerka instability, resulting in a fractal morphology with many irregular branches. Thus, the crystal morphology of the obstructed atomic insulator differs significantly from that of the normal insulator, even in the case $e^{(\textrm{n})}_b=e_b^{(\textrm{OAI})}$.

\section{Comparison at the same $\Delta \mu$}

\begin{figure}
    \includegraphics[width=1.\columnwidth]{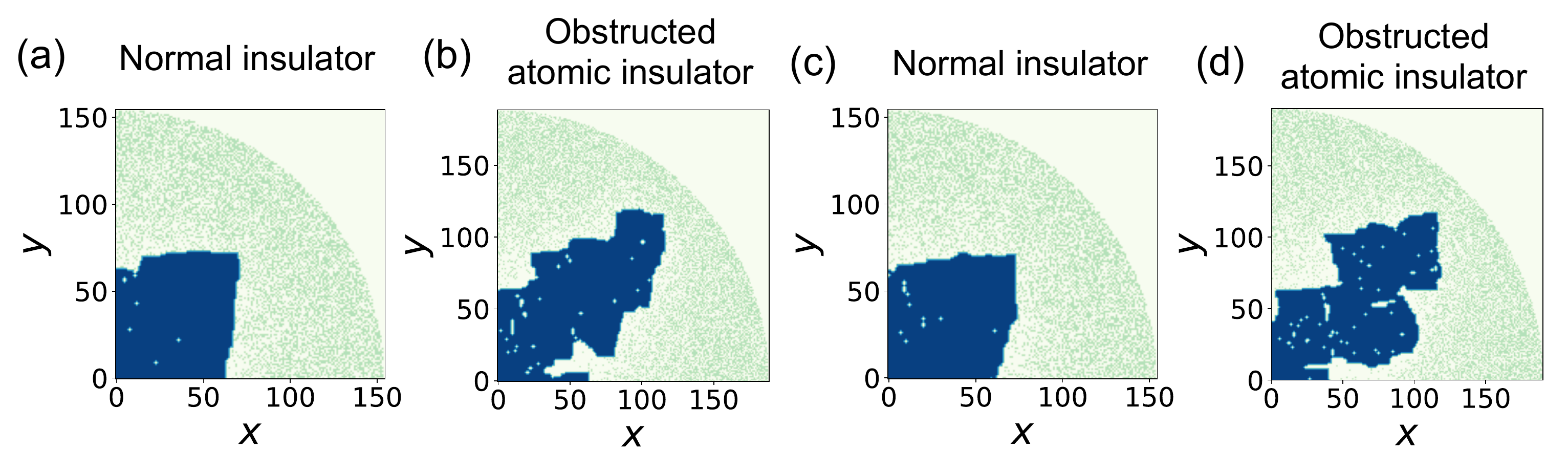}
	\caption{(a)-(d)~Crystal shapes obtained from the simulations for (a,b) $\Delta \mu/k_{\rm B}T = -9$ and (c,d) $\Delta \mu/k_{\rm B}T = -9.25$. 
    The blue, light blue, and light green sites indicate the solid atoms, interfacial solid atoms, and gas atoms, respectively. 
    The parameters are set to (i)~$e_b=0.6439+0.2413+0.1488$, $e_s = 0$, $e_t = 0$ for the normal insulator~[(a) and (c)] and (ii)~$e_b=0.6439$, $e_s = 0.2413$, $e_t = 0.1488$ for the obstructed atomic insulator~[(b) and (d)]. 
    We choose $k_{\rm B}T=0.16$ in both the (i) and (ii) cases. All the crystal shapes are obtained with $t=1200$.
    }
	\label{fig:comp_same_mu}
\end{figure}

In this section, we have compared the crystal morphology of the obstructed atomic insulator to that of the normal insulator at the same $\Delta \mu$. 
Figure~\ref{fig:comp_same_mu} shows the crystal shapes obtained from the simulations for the normal insulator~[Figs.~\ref{fig:comp_same_mu}(a) and~\ref{fig:comp_same_mu}(c)] and the obstructed atomic insulator~[Figs.~\ref{fig:comp_same_mu}(b) and~\ref{fig:comp_same_mu}(d)] at identical values of $\Delta \mu$.
The results in Figs.~\ref{fig:comp_same_mu}(b) and~\ref{fig:comp_same_mu}(d) fall into the OAI-dominant region in Fig.~6(c) ($9 k_{\rm B}T\leq |\Delta \mu|\leq 9.25k_{\rm B}T$), where our key findings are demonstrated.
These results indicate that the obstructed atomic insulator exhibits morphologies with more developed corners compared to the normal insulator.
This morphological difference between these two phases is quantified by the fractal dimension $D_f$.
The fractal dimensions are found to be $D_f=1.74$ for Fig.~\ref{fig:comp_same_mu}(b) and $D_f=1.73$ for Fig.~\ref{fig:comp_same_mu}(d) in the obstructed atomic insulator phase, whereas they are $D_f=1.90$ for Fig.~\ref{fig:comp_same_mu}(a) and $D_f=1.90$ for Fig.~\ref{fig:comp_same_mu}(c) in the normal insulator phase. 
These values of $D_f$ are calculated for the single realizations in Fig.~\ref{fig:comp_same_mu}.
Furthermore, as shown in Fig.~2(c), the ensemble-averaged $D_f$ of the obstructed atomic insulator is consistently lower than that of the normal insulator at the same $|\Delta \mu|$. Thus, even when we compare the results in both phases at the same $|\Delta \mu|$, the crystal morphology of the obstructed atomic insulator is distinguishable from that of the normal insulator.

\begin{thebibliography}{60}%
\makeatletter
\providecommand \@ifxundefined [1]{%
 \@ifx{#1\undefined}
}%
\providecommand \@ifnum [1]{%
 \ifnum #1\expandafter \@firstoftwo
 \else \expandafter \@secondoftwo
 \fi
}%
\providecommand \@ifx [1]{%
 \ifx #1\expandafter \@firstoftwo
 \else \expandafter \@secondoftwo
 \fi
}%
\providecommand \natexlab [1]{#1}%
\providecommand \enquote  [1]{``#1''}%
\providecommand \bibnamefont  [1]{#1}%
\providecommand \bibfnamefont [1]{#1}%
\providecommand \citenamefont [1]{#1}%
\providecommand \href@noop [0]{\@secondoftwo}%
\providecommand \href [0]{\begingroup \@sanitize@url \@href}%
\providecommand \@href[1]{\@@startlink{#1}\@@href}%
\providecommand \@@href[1]{\endgroup#1\@@endlink}%
\providecommand \@sanitize@url [0]{\catcode `\\12\catcode `\$12\catcode
  `\&12\catcode `\#12\catcode `\^12\catcode `\_12\catcode `\%12\relax}%
\providecommand \@@startlink[1]{}%
\providecommand \@@endlink[0]{}%
\providecommand \url  [0]{\begingroup\@sanitize@url \@url }%
\providecommand \@url [1]{\endgroup\@href {#1}{\urlprefix }}%
\providecommand \urlprefix  [0]{URL }%
\providecommand \Eprint [0]{\href }%
\providecommand \doibase [0]{https://doi.org/}%
\providecommand \selectlanguage [0]{\@gobble}%
\providecommand \bibinfo  [0]{\@secondoftwo}%
\providecommand \bibfield  [0]{\@secondoftwo}%
\providecommand \translation [1]{[#1]}%
\providecommand \BibitemOpen [0]{}%
\providecommand \bibitemStop [0]{}%
\providecommand \bibitemNoStop [0]{.\EOS\space}%
\providecommand \EOS [0]{\spacefactor3000\relax}%
\providecommand \BibitemShut  [1]{\csname bibitem#1\endcsname}%
\let\auto@bib@innerbib\@empty
\bibitem [{\citenamefont {Langer}(1980)}]{RevModPhys.52.1}%
  \BibitemOpen
  \bibfield  {author} {\bibinfo {author} {\bibfnamefont {J.~S.}\ \bibnamefont
  {Langer}},\ }\bibfield  {title} {\bibinfo {title} {Instabilities and pattern
  formation in crystal growth},\ }\href
  {https://doi.org/10.1103/RevModPhys.52.1} {\bibfield  {journal} {\bibinfo
  {journal} {Rev. Mod. Phys.}\ }\textbf {\bibinfo {volume} {52}},\ \bibinfo
  {pages} {1} (\bibinfo {year} {1980})}\BibitemShut {NoStop}%
\bibitem [{\citenamefont {Ben-Jacob}\ and\ \citenamefont
  {Garik}(1990)}]{ben1990formation}%
  \BibitemOpen
  \bibfield  {author} {\bibinfo {author} {\bibfnamefont {E.}~\bibnamefont
  {Ben-Jacob}}\ and\ \bibinfo {author} {\bibfnamefont {P.}~\bibnamefont
  {Garik}},\ }\bibfield  {title} {\bibinfo {title} {{The formation of patterns
  in non-equilibrium growth}},\ }\href
  {https://www.nature.com/articles/343523a0} {\bibfield  {journal} {\bibinfo
  {journal} {Nature}\ }\textbf {\bibinfo {volume} {343}},\ \bibinfo {pages}
  {523} (\bibinfo {year} {1990})}\BibitemShut {NoStop}%
\bibitem [{\citenamefont {Amelinckx}(1953)}]{Amelinckx01031953}%
  \BibitemOpen
  \bibfield  {author} {\bibinfo {author} {\bibfnamefont {S.}~\bibnamefont
  {Amelinckx}},\ }\bibfield  {title} {\bibinfo {title} {{XXXVII. A dislocation
  mechanism for the growth of hopper crystal faces and the growth of salol
  crystals from solution and from the melt}},\ }\href
  {https://doi.org/10.1080/14786440308520314} {\bibfield  {journal} {\bibinfo
  {journal} {Lond. Edinb. Dubl. Philos. Mag. J. Sci.}\ }\textbf {\bibinfo
  {volume} {44}},\ \bibinfo {pages} {337} (\bibinfo {year} {1953})}\BibitemShut
  {NoStop}%
\bibitem [{\citenamefont {Sunagawa}(1999)}]{Sunagawa1999}%
  \BibitemOpen
  \bibfield  {author} {\bibinfo {author} {\bibfnamefont {I.}~\bibnamefont
  {Sunagawa}},\ }\bibfield  {title} {\bibinfo {title} {{Growth and Morphology
  of Crystals}},\ }\href@noop {} {\bibfield  {journal} {\bibinfo  {journal}
  {Forma}\ }\textbf {\bibinfo {volume} {14}},\ \bibinfo {pages} {147} (\bibinfo
  {year} {1999})}\BibitemShut {NoStop}%
\bibitem [{\citenamefont {Fontana}\ \emph {et~al.}(2011)\citenamefont
  {Fontana}, \citenamefont {Schefer},\ and\ \citenamefont
  {Pettit}}]{FONTANA2011207}%
  \BibitemOpen
  \bibfield  {author} {\bibinfo {author} {\bibfnamefont {P.}~\bibnamefont
  {Fontana}}, \bibinfo {author} {\bibfnamefont {J.}~\bibnamefont {Schefer}},\
  and\ \bibinfo {author} {\bibfnamefont {D.}~\bibnamefont {Pettit}},\
  }\bibfield  {title} {\bibinfo {title} {{Characterization of sodium chloride
  crystals grown in microgravity}},\ }\href
  {https://doi.org/https://doi.org/10.1016/j.jcrysgro.2011.04.001} {\bibfield
  {journal} {\bibinfo  {journal} {J. Cryst. Growth}\ }\textbf {\bibinfo
  {volume} {324}},\ \bibinfo {pages} {207} (\bibinfo {year}
  {2011})}\BibitemShut {NoStop}%
\bibitem [{\citenamefont {Yin}\ \emph {et~al.}(2014)\citenamefont {Yin},
  \citenamefont {Wang},\ and\ \citenamefont {Chen}}]{YIN2014131}%
  \BibitemOpen
  \bibfield  {author} {\bibinfo {author} {\bibfnamefont {H.}~\bibnamefont
  {Yin}}, \bibinfo {author} {\bibfnamefont {Q.}~\bibnamefont {Wang}},\ and\
  \bibinfo {author} {\bibfnamefont {G.}~\bibnamefont {Chen}},\ }\bibfield
  {title} {\bibinfo {title} {{Probing the growth mechanism of PbTe hopper-like
  crystal and ultra-long nanowires with rough surface synthesized through
  acetone-assisted solvothermal method}},\ }\href
  {https://doi.org/https://doi.org/10.1016/j.cej.2013.09.078} {\bibfield
  {journal} {\bibinfo  {journal} {Chem. Eng. J.}\ }\textbf {\bibinfo {volume}
  {236}},\ \bibinfo {pages} {131} (\bibinfo {year} {2014})}\BibitemShut
  {NoStop}%
\bibitem [{\citenamefont {Desarnaud}\ \emph {et~al.}(2018)\citenamefont
  {Desarnaud}, \citenamefont {Derluyn}, \citenamefont {Carmeliet},
  \citenamefont {Bonn},\ and\ \citenamefont {Shahidzadeh}}]{Desarnaud2018}%
  \BibitemOpen
  \bibfield  {author} {\bibinfo {author} {\bibfnamefont {J.}~\bibnamefont
  {Desarnaud}}, \bibinfo {author} {\bibfnamefont {H.}~\bibnamefont {Derluyn}},
  \bibinfo {author} {\bibfnamefont {J.}~\bibnamefont {Carmeliet}}, \bibinfo
  {author} {\bibfnamefont {D.}~\bibnamefont {Bonn}},\ and\ \bibinfo {author}
  {\bibfnamefont {N.}~\bibnamefont {Shahidzadeh}},\ }\bibfield  {title}
  {\bibinfo {title} {{Hopper Growth of Salt Crystals}},\ }\href
  {https://doi.org/10.1021/acs.jpclett.8b01082} {\bibfield  {journal} {\bibinfo
   {journal} {J. Phys. Chem. Lett.}\ }\textbf {\bibinfo {volume} {9}},\
  \bibinfo {pages} {2961} (\bibinfo {year} {2018})}\BibitemShut {NoStop}%
\bibitem [{\citenamefont {Pettit}\ and\ \citenamefont
  {Fontana}(2019)}]{Pettit2019}%
  \BibitemOpen
  \bibfield  {author} {\bibinfo {author} {\bibfnamefont {D.}~\bibnamefont
  {Pettit}}\ and\ \bibinfo {author} {\bibfnamefont {P.}~\bibnamefont
  {Fontana}},\ }\bibfield  {title} {\bibinfo {title} {{Comparison of sodium
  chloride hopper cubes grown under microgravity and terrestrial conditions}},\
  }\href {https://doi.org/10.1038/s41526-019-0085-0} {\bibfield  {journal}
  {\bibinfo  {journal} {npj Microgravity}\ }\textbf {\bibinfo {volume} {5}},\
  \bibinfo {pages} {25} (\bibinfo {year} {2019})}\BibitemShut {NoStop}%
\bibitem [{\citenamefont {Yang}\ \emph {et~al.}(2020)\citenamefont {Yang},
  \citenamefont {Zhang}, \citenamefont {Zhang}, \citenamefont {Fu},
  \citenamefont {Tao}, \citenamefont {Song}, \citenamefont {Shang},\ and\
  \citenamefont {Deng}}]{https://doi.org/10.1002/adfm.201908108}%
  \BibitemOpen
  \bibfield  {author} {\bibinfo {author} {\bibfnamefont {Z.}~\bibnamefont
  {Yang}}, \bibinfo {author} {\bibfnamefont {J.}~\bibnamefont {Zhang}},
  \bibinfo {author} {\bibfnamefont {L.}~\bibnamefont {Zhang}}, \bibinfo
  {author} {\bibfnamefont {B.}~\bibnamefont {Fu}}, \bibinfo {author}
  {\bibfnamefont {P.}~\bibnamefont {Tao}}, \bibinfo {author} {\bibfnamefont
  {C.}~\bibnamefont {Song}}, \bibinfo {author} {\bibfnamefont {W.}~\bibnamefont
  {Shang}},\ and\ \bibinfo {author} {\bibfnamefont {T.}~\bibnamefont {Deng}},\
  }\bibfield  {title} {\bibinfo {title} {{Self-Assembly in Hopper-Shaped
  Crystals}},\ }\href {https://doi.org/https://doi.org/10.1002/adfm.201908108}
  {\bibfield  {journal} {\bibinfo  {journal} {Adv. Funct. Mater.}\ }\textbf
  {\bibinfo {volume} {30}},\ \bibinfo {pages} {1908108} (\bibinfo {year}
  {2020})}\BibitemShut {NoStop}%
\bibitem [{\citenamefont {Bollada}\ \emph {et~al.}(2023)\citenamefont
  {Bollada}, \citenamefont {Jimack},\ and\ \citenamefont
  {Mullis}}]{Bollada2023}%
  \BibitemOpen
  \bibfield  {author} {\bibinfo {author} {\bibfnamefont {P.~C.}\ \bibnamefont
  {Bollada}}, \bibinfo {author} {\bibfnamefont {P.~K.}\ \bibnamefont
  {Jimack}},\ and\ \bibinfo {author} {\bibfnamefont {A.~M.}\ \bibnamefont
  {Mullis}},\ }\bibfield  {title} {\bibinfo {title} {{Phase field modelling of
  hopper crystal growth in alloys}},\ }\href
  {https://doi.org/10.1038/s41598-023-38741-2} {\bibfield  {journal} {\bibinfo
  {journal} {Sci. Rep.}\ }\textbf {\bibinfo {volume} {13}},\ \bibinfo {pages}
  {12637} (\bibinfo {year} {2023})}\BibitemShut {NoStop}%
\bibitem [{\citenamefont {Pimpinelli}\ and\ \citenamefont
  {Villain}(1998)}]{Pimpinelli_Villain_1998}%
  \BibitemOpen
  \bibfield  {author} {\bibinfo {author} {\bibfnamefont {A.}~\bibnamefont
  {Pimpinelli}}\ and\ \bibinfo {author} {\bibfnamefont {J.}~\bibnamefont
  {Villain}},\ }\href@noop {} {\emph {\bibinfo {title} {{Physics of Crystal
  Growth}}}}\ (\bibinfo  {publisher} {Cambridge University Press},\ \bibinfo
  {year} {1998})\BibitemShut {NoStop}%
\bibitem [{\citenamefont {Sunagawa}(2007)}]{sunagawa2007crystals}%
  \BibitemOpen
  \bibfield  {author} {\bibinfo {author} {\bibfnamefont {I.}~\bibnamefont
  {Sunagawa}},\ }\href@noop {} {\emph {\bibinfo {title} {Crystals: growth,
  morphology, $\&$ perfection}}}\ (\bibinfo  {publisher} {Cambridge University
  Press},\ \bibinfo {year} {2007})\BibitemShut {NoStop}%
\bibitem [{\citenamefont {Hasan}\ and\ \citenamefont
  {Kane}(2010)}]{RevModPhys.82.3045}%
  \BibitemOpen
  \bibfield  {author} {\bibinfo {author} {\bibfnamefont {M.~Z.}\ \bibnamefont
  {Hasan}}\ and\ \bibinfo {author} {\bibfnamefont {C.~L.}\ \bibnamefont
  {Kane}},\ }\bibfield  {title} {\bibinfo {title} {{Colloquium: Topological
  insulators}},\ }\href {https://doi.org/10.1103/RevModPhys.82.3045} {\bibfield
   {journal} {\bibinfo  {journal} {Rev. Mod. Phys.}\ }\textbf {\bibinfo
  {volume} {82}},\ \bibinfo {pages} {3045} (\bibinfo {year}
  {2010})}\BibitemShut {NoStop}%
\bibitem [{\citenamefont {Qi}\ and\ \citenamefont
  {Zhang}(2011)}]{RevModPhys.83.1057}%
  \BibitemOpen
  \bibfield  {author} {\bibinfo {author} {\bibfnamefont {X.-L.}\ \bibnamefont
  {Qi}}\ and\ \bibinfo {author} {\bibfnamefont {S.-C.}\ \bibnamefont {Zhang}},\
  }\bibfield  {title} {\bibinfo {title} {{Topological insulators and
  superconductors}},\ }\href {https://doi.org/10.1103/RevModPhys.83.1057}
  {\bibfield  {journal} {\bibinfo  {journal} {Rev. Mod. Phys.}\ }\textbf
  {\bibinfo {volume} {83}},\ \bibinfo {pages} {1057} (\bibinfo {year}
  {2011})}\BibitemShut {NoStop}%
\bibitem [{\citenamefont {Fu}(2011)}]{PhysRevLett.106.106802}%
  \BibitemOpen
  \bibfield  {author} {\bibinfo {author} {\bibfnamefont {L.}~\bibnamefont
  {Fu}},\ }\bibfield  {title} {\bibinfo {title} {Topological crystalline
  insulators},\ }\href {https://doi.org/10.1103/PhysRevLett.106.106802}
  {\bibfield  {journal} {\bibinfo  {journal} {Phys. Rev. Lett.}\ }\textbf
  {\bibinfo {volume} {106}},\ \bibinfo {pages} {106802} (\bibinfo {year}
  {2011})}\BibitemShut {NoStop}%
\bibitem [{\citenamefont {Hsieh}\ \emph {et~al.}(2012)\citenamefont {Hsieh},
  \citenamefont {Lin}, \citenamefont {Liu}, \citenamefont {Duan}, \citenamefont
  {Bansil},\ and\ \citenamefont {Fu}}]{hsieh2012topological}%
  \BibitemOpen
  \bibfield  {author} {\bibinfo {author} {\bibfnamefont {T.~H.}\ \bibnamefont
  {Hsieh}}, \bibinfo {author} {\bibfnamefont {H.}~\bibnamefont {Lin}}, \bibinfo
  {author} {\bibfnamefont {J.}~\bibnamefont {Liu}}, \bibinfo {author}
  {\bibfnamefont {W.}~\bibnamefont {Duan}}, \bibinfo {author} {\bibfnamefont
  {A.}~\bibnamefont {Bansil}},\ and\ \bibinfo {author} {\bibfnamefont
  {L.}~\bibnamefont {Fu}},\ }\bibfield  {title} {\bibinfo {title} {{Topological
  crystalline insulators in the SnTe material class}},\ }\href
  {https://www.nature.com/articles/ncomms1969} {\bibfield  {journal} {\bibinfo
  {journal} {Nat. Commun.}\ }\textbf {\bibinfo {volume} {3}},\ \bibinfo {pages}
  {982} (\bibinfo {year} {2012})}\BibitemShut {NoStop}%
\bibitem [{\citenamefont {Slager}\ \emph {et~al.}(2013)\citenamefont {Slager},
  \citenamefont {Mesaros}, \citenamefont {Juri{\v c}i{\'c}},\ and\
  \citenamefont {Zaanen}}]{slager2013space}%
  \BibitemOpen
  \bibfield  {author} {\bibinfo {author} {\bibfnamefont {R.-J.}\ \bibnamefont
  {Slager}}, \bibinfo {author} {\bibfnamefont {A.}~\bibnamefont {Mesaros}},
  \bibinfo {author} {\bibfnamefont {V.}~\bibnamefont {Juri{\v c}i{\'c}}},\ and\
  \bibinfo {author} {\bibfnamefont {J.}~\bibnamefont {Zaanen}},\ }\bibfield
  {title} {\bibinfo {title} {The space group classification of topological
  band-insulators},\ }\href {https://doi.org/10.1038/nphys2513} {\bibfield
  {journal} {\bibinfo  {journal} {Nat. Phys.}\ }\textbf {\bibinfo {volume}
  {9}},\ \bibinfo {pages} {98} (\bibinfo {year} {2013})}\BibitemShut {NoStop}%
\bibitem [{\citenamefont {Sitte}\ \emph {et~al.}(2012)\citenamefont {Sitte},
  \citenamefont {Rosch}, \citenamefont {Altman},\ and\ \citenamefont
  {Fritz}}]{PhysRevLett.108.126807}%
  \BibitemOpen
  \bibfield  {author} {\bibinfo {author} {\bibfnamefont {M.}~\bibnamefont
  {Sitte}}, \bibinfo {author} {\bibfnamefont {A.}~\bibnamefont {Rosch}},
  \bibinfo {author} {\bibfnamefont {E.}~\bibnamefont {Altman}},\ and\ \bibinfo
  {author} {\bibfnamefont {L.}~\bibnamefont {Fritz}},\ }\bibfield  {title}
  {\bibinfo {title} {{Topological Insulators in Magnetic Fields: Quantum Hall
  Effect and Edge Channels with a Nonquantized $\ensuremath{\theta}$ Term}},\
  }\href {https://doi.org/10.1103/PhysRevLett.108.126807} {\bibfield  {journal}
  {\bibinfo  {journal} {Phys. Rev. Lett.}\ }\textbf {\bibinfo {volume} {108}},\
  \bibinfo {pages} {126807} (\bibinfo {year} {2012})}\BibitemShut {NoStop}%
\bibitem [{\citenamefont {Zhang}\ \emph {et~al.}(2013)\citenamefont {Zhang},
  \citenamefont {Kane},\ and\ \citenamefont {Mele}}]{PhysRevLett.110.046404}%
  \BibitemOpen
  \bibfield  {author} {\bibinfo {author} {\bibfnamefont {F.}~\bibnamefont
  {Zhang}}, \bibinfo {author} {\bibfnamefont {C.~L.}\ \bibnamefont {Kane}},\
  and\ \bibinfo {author} {\bibfnamefont {E.~J.}\ \bibnamefont {Mele}},\
  }\bibfield  {title} {\bibinfo {title} {{Surface State Magnetization and
  Chiral Edge States on Topological Insulators}},\ }\href
  {https://doi.org/10.1103/PhysRevLett.110.046404} {\bibfield  {journal}
  {\bibinfo  {journal} {Phys. Rev. Lett.}\ }\textbf {\bibinfo {volume} {110}},\
  \bibinfo {pages} {046404} (\bibinfo {year} {2013})}\BibitemShut {NoStop}%
\bibitem [{\citenamefont {Benalcazar}\ \emph
  {et~al.}(2017{\natexlab{a}})\citenamefont {Benalcazar}, \citenamefont
  {Bernevig},\ and\ \citenamefont {Hughes}}]{benalcazar2017quantized}%
  \BibitemOpen
  \bibfield  {author} {\bibinfo {author} {\bibfnamefont {W.~A.}\ \bibnamefont
  {Benalcazar}}, \bibinfo {author} {\bibfnamefont {B.~A.}\ \bibnamefont
  {Bernevig}},\ and\ \bibinfo {author} {\bibfnamefont {T.~L.}\ \bibnamefont
  {Hughes}},\ }\bibfield  {title} {\bibinfo {title} {{Quantized electric
  multipole insulators}},\ }\href {https://doi.org/10.1126/science.aah6442}
  {\bibfield  {journal} {\bibinfo  {journal} {Science}\ }\textbf {\bibinfo
  {volume} {357}},\ \bibinfo {pages} {61} (\bibinfo {year}
  {2017}{\natexlab{a}})}\BibitemShut {NoStop}%
\bibitem [{\citenamefont {Benalcazar}\ \emph
  {et~al.}(2017{\natexlab{b}})\citenamefont {Benalcazar}, \citenamefont
  {Bernevig},\ and\ \citenamefont {Hughes}}]{PhysRevB.96.245115}%
  \BibitemOpen
  \bibfield  {author} {\bibinfo {author} {\bibfnamefont {W.~A.}\ \bibnamefont
  {Benalcazar}}, \bibinfo {author} {\bibfnamefont {B.~A.}\ \bibnamefont
  {Bernevig}},\ and\ \bibinfo {author} {\bibfnamefont {T.~L.}\ \bibnamefont
  {Hughes}},\ }\bibfield  {title} {\bibinfo {title} {{Electric multipole
  moments, topological multipole moment pumping, and chiral hinge states in
  crystalline insulators}},\ }\href
  {https://doi.org/10.1103/PhysRevB.96.245115} {\bibfield  {journal} {\bibinfo
  {journal} {Phys. Rev. B}\ }\textbf {\bibinfo {volume} {96}},\ \bibinfo
  {pages} {245115} (\bibinfo {year} {2017}{\natexlab{b}})}\BibitemShut
  {NoStop}%
\bibitem [{\citenamefont {Langbehn}\ \emph {et~al.}(2017)\citenamefont
  {Langbehn}, \citenamefont {Peng}, \citenamefont {Trifunovic}, \citenamefont
  {von Oppen},\ and\ \citenamefont {Brouwer}}]{PhysRevLett.119.246401}%
  \BibitemOpen
  \bibfield  {author} {\bibinfo {author} {\bibfnamefont {J.}~\bibnamefont
  {Langbehn}}, \bibinfo {author} {\bibfnamefont {Y.}~\bibnamefont {Peng}},
  \bibinfo {author} {\bibfnamefont {L.}~\bibnamefont {Trifunovic}}, \bibinfo
  {author} {\bibfnamefont {F.}~\bibnamefont {von Oppen}},\ and\ \bibinfo
  {author} {\bibfnamefont {P.~W.}\ \bibnamefont {Brouwer}},\ }\bibfield
  {title} {\bibinfo {title} {{Reflection-Symmetric Second-Order Topological
  Insulators and Superconductors}},\ }\href
  {https://doi.org/10.1103/PhysRevLett.119.246401} {\bibfield  {journal}
  {\bibinfo  {journal} {Phys. Rev. Lett.}\ }\textbf {\bibinfo {volume} {119}},\
  \bibinfo {pages} {246401} (\bibinfo {year} {2017})}\BibitemShut {NoStop}%
\bibitem [{\citenamefont {Song}\ \emph {et~al.}(2017)\citenamefont {Song},
  \citenamefont {Fang},\ and\ \citenamefont {Fang}}]{PhysRevLett.119.246402}%
  \BibitemOpen
  \bibfield  {author} {\bibinfo {author} {\bibfnamefont {Z.}~\bibnamefont
  {Song}}, \bibinfo {author} {\bibfnamefont {Z.}~\bibnamefont {Fang}},\ and\
  \bibinfo {author} {\bibfnamefont {C.}~\bibnamefont {Fang}},\ }\bibfield
  {title} {\bibinfo {title} {{$(d\ensuremath{-}2)$-Dimensional Edge States of
  Rotation Symmetry Protected Topological States}},\ }\href
  {https://doi.org/10.1103/PhysRevLett.119.246402} {\bibfield  {journal}
  {\bibinfo  {journal} {Phys. Rev. Lett.}\ }\textbf {\bibinfo {volume} {119}},\
  \bibinfo {pages} {246402} (\bibinfo {year} {2017})}\BibitemShut {NoStop}%
\bibitem [{\citenamefont {Schindler}\ \emph
  {et~al.}(2018{\natexlab{a}})\citenamefont {Schindler}, \citenamefont {Cook},
  \citenamefont {Vergniory}, \citenamefont {Wang}, \citenamefont {Parkin},
  \citenamefont {Bernevig},\ and\ \citenamefont
  {Neupert}}]{schindler2018higher}%
  \BibitemOpen
  \bibfield  {author} {\bibinfo {author} {\bibfnamefont {F.}~\bibnamefont
  {Schindler}}, \bibinfo {author} {\bibfnamefont {A.~M.}\ \bibnamefont {Cook}},
  \bibinfo {author} {\bibfnamefont {M.~G.}\ \bibnamefont {Vergniory}}, \bibinfo
  {author} {\bibfnamefont {Z.}~\bibnamefont {Wang}}, \bibinfo {author}
  {\bibfnamefont {S.~S.}\ \bibnamefont {Parkin}}, \bibinfo {author}
  {\bibfnamefont {B.~A.}\ \bibnamefont {Bernevig}},\ and\ \bibinfo {author}
  {\bibfnamefont {T.}~\bibnamefont {Neupert}},\ }\bibfield  {title} {\bibinfo
  {title} {{Higher-order topological insulators}},\ }\href
  {https://doi.org/10.1126/sciadv.aat0346} {\bibfield  {journal} {\bibinfo
  {journal} {Sci. Adv.}\ }\textbf {\bibinfo {volume} {4}},\ \bibinfo {pages}
  {eaat0346} (\bibinfo {year} {2018}{\natexlab{a}})}\BibitemShut {NoStop}%
\bibitem [{\citenamefont {Tanaka}\ \emph {et~al.}(2022)\citenamefont {Tanaka},
  \citenamefont {Zhang}, \citenamefont {Uwaha},\ and\ \citenamefont
  {Murakami}}]{PhysRevLett.129.046802}%
  \BibitemOpen
  \bibfield  {author} {\bibinfo {author} {\bibfnamefont {Y.}~\bibnamefont
  {Tanaka}}, \bibinfo {author} {\bibfnamefont {T.}~\bibnamefont {Zhang}},
  \bibinfo {author} {\bibfnamefont {M.}~\bibnamefont {Uwaha}},\ and\ \bibinfo
  {author} {\bibfnamefont {S.}~\bibnamefont {Murakami}},\ }\bibfield  {title}
  {\bibinfo {title} {{Anomalous Crystal Shapes of Topological Crystalline
  Insulators}},\ }\href {https://doi.org/10.1103/PhysRevLett.129.046802}
  {\bibfield  {journal} {\bibinfo  {journal} {Phys. Rev. Lett.}\ }\textbf
  {\bibinfo {volume} {129}},\ \bibinfo {pages} {046802} (\bibinfo {year}
  {2022})}\BibitemShut {NoStop}%
\bibitem [{\citenamefont {Tanaka}\ and\ \citenamefont
  {Murakami}(2023)}]{PhysRevB.107.245148}%
  \BibitemOpen
  \bibfield  {author} {\bibinfo {author} {\bibfnamefont {Y.}~\bibnamefont
  {Tanaka}}\ and\ \bibinfo {author} {\bibfnamefont {S.}~\bibnamefont
  {Murakami}},\ }\bibfield  {title} {\bibinfo {title} {Effects of first- and
  second-order topological phases on equilibrium crystal shapes},\ }\href
  {https://doi.org/10.1103/PhysRevB.107.245148} {\bibfield  {journal} {\bibinfo
   {journal} {Phys. Rev. B}\ }\textbf {\bibinfo {volume} {107}},\ \bibinfo
  {pages} {245148} (\bibinfo {year} {2023})}\BibitemShut {NoStop}%
\bibitem [{\citenamefont {Mondal}\ and\ \citenamefont
  {Agarwala}(2025)}]{Mondal2025}%
  \BibitemOpen
  \bibfield  {author} {\bibinfo {author} {\bibfnamefont {S.}~\bibnamefont
  {Mondal}}\ and\ \bibinfo {author} {\bibfnamefont {A.}~\bibnamefont
  {Agarwala}},\ }\bibfield  {title} {\bibinfo {title} {{Interfacial line energy
  of a topological phase}},\ }\href {https://doi.org/10.1103/zndb-hzkk}
  {\bibfield  {journal} {\bibinfo  {journal} {Phys. Rev. B}\ }\textbf {\bibinfo
  {volume} {112}},\ \bibinfo {pages} {245159} (\bibinfo {year}
  {2025})}\BibitemShut {NoStop}%
\bibitem [{\citenamefont {Schindler}\ \emph
  {et~al.}(2018{\natexlab{b}})\citenamefont {Schindler}, \citenamefont {Wang},
  \citenamefont {Vergniory}, \citenamefont {Cook}, \citenamefont {Murani},
  \citenamefont {Sengupta}, \citenamefont {Kasumov}, \citenamefont {Deblock},
  \citenamefont {Jeon}, \citenamefont {Drozdov}, \citenamefont {Bouchiat},
  \citenamefont {Gu\'eron}, \citenamefont {Yazdani}, \citenamefont {Bernevig},\
  and\ \citenamefont {Neupert}}]{schindler2018higherbismuth}%
  \BibitemOpen
  \bibfield  {author} {\bibinfo {author} {\bibfnamefont {F.}~\bibnamefont
  {Schindler}}, \bibinfo {author} {\bibfnamefont {Z.}~\bibnamefont {Wang}},
  \bibinfo {author} {\bibfnamefont {M.~G.}\ \bibnamefont {Vergniory}}, \bibinfo
  {author} {\bibfnamefont {A.~M.}\ \bibnamefont {Cook}}, \bibinfo {author}
  {\bibfnamefont {A.}~\bibnamefont {Murani}}, \bibinfo {author} {\bibfnamefont
  {S.}~\bibnamefont {Sengupta}}, \bibinfo {author} {\bibfnamefont {A.~Y.}\
  \bibnamefont {Kasumov}}, \bibinfo {author} {\bibfnamefont {R.}~\bibnamefont
  {Deblock}}, \bibinfo {author} {\bibfnamefont {S.}~\bibnamefont {Jeon}},
  \bibinfo {author} {\bibfnamefont {I.}~\bibnamefont {Drozdov}}, \bibinfo
  {author} {\bibfnamefont {H.}~\bibnamefont {Bouchiat}}, \bibinfo {author}
  {\bibfnamefont {S.}~\bibnamefont {Gu\'eron}}, \bibinfo {author}
  {\bibfnamefont {A.}~\bibnamefont {Yazdani}}, \bibinfo {author} {\bibfnamefont
  {B.~A.}\ \bibnamefont {Bernevig}},\ and\ \bibinfo {author} {\bibfnamefont
  {T.}~\bibnamefont {Neupert}},\ }\bibfield  {title} {\bibinfo {title}
  {{Higher-order topology in bismuth}},\ }\href
  {https://doi.org/10.1038/s41567-018-0224-7} {\bibfield  {journal} {\bibinfo
  {journal} {Nat. Phys.}\ }\textbf {\bibinfo {volume} {14}},\ \bibinfo {pages}
  {918} (\bibinfo {year} {2018}{\natexlab{b}})}\BibitemShut {NoStop}%
\bibitem [{\citenamefont {Robredo}\ \emph {et~al.}(2019)\citenamefont
  {Robredo}, \citenamefont {Vergniory},\ and\ \citenamefont
  {Bradlyn}}]{Robredo2019}%
  \BibitemOpen
  \bibfield  {author} {\bibinfo {author} {\bibfnamefont {I.}~\bibnamefont
  {Robredo}}, \bibinfo {author} {\bibfnamefont {M.~G.}\ \bibnamefont
  {Vergniory}},\ and\ \bibinfo {author} {\bibfnamefont {B.}~\bibnamefont
  {Bradlyn}},\ }\bibfield  {title} {\bibinfo {title} {{Higher-order and
  crystalline topology in a phenomenological tight-binding model of lead
  telluride}},\ }\href {https://doi.org/10.1103/PhysRevMaterials.3.041202}
  {\bibfield  {journal} {\bibinfo  {journal} {Phys. Rev. Mater.}\ }\textbf
  {\bibinfo {volume} {3}},\ \bibinfo {pages} {041202} (\bibinfo {year}
  {2019})}\BibitemShut {NoStop}%
\bibitem [{\citenamefont {Watanabe}\ and\ \citenamefont
  {Po}(2021)}]{Watanabe2021}%
  \BibitemOpen
  \bibfield  {author} {\bibinfo {author} {\bibfnamefont {H.}~\bibnamefont
  {Watanabe}}\ and\ \bibinfo {author} {\bibfnamefont {H.~C.}\ \bibnamefont
  {Po}},\ }\bibfield  {title} {\bibinfo {title} {{Fractional Corner Charge of
  Sodium Chloride}},\ }\href {https://doi.org/10.1103/PhysRevX.11.041064}
  {\bibfield  {journal} {\bibinfo  {journal} {Phys. Rev. X}\ }\textbf {\bibinfo
  {volume} {11}},\ \bibinfo {pages} {041064} (\bibinfo {year}
  {2021})}\BibitemShut {NoStop}%
\bibitem [{\citenamefont {Hausdorff}(1919)}]{Hausdorff1919}%
  \BibitemOpen
  \bibfield  {author} {\bibinfo {author} {\bibfnamefont {F.}~\bibnamefont
  {Hausdorff}},\ }\bibfield  {title} {\bibinfo {title} {{Dimension und
  äußeres Maß}},\ }\href {http://eudml.org/doc/158784} {\bibfield  {journal}
  {\bibinfo  {journal} {Mathematische Annalen}\ }\textbf {\bibinfo {volume}
  {79}},\ \bibinfo {pages} {157} (\bibinfo {year} {1919})}\BibitemShut
  {NoStop}%
\bibitem [{\citenamefont {Mandelbrot}(1982)}]{Manderlbrot_fractal_nature}%
  \BibitemOpen
  \bibfield  {author} {\bibinfo {author} {\bibfnamefont {B.~B.}\ \bibnamefont
  {Mandelbrot}},\ }\href@noop {} {\emph {\bibinfo {title} {{The Fractal
  Geometry of Nature}}}}\ (\bibinfo  {publisher} {Freeman},\ \bibinfo {address}
  {San Francisco},\ \bibinfo {year} {1982})\BibitemShut {NoStop}%
\bibitem [{\citenamefont
  {Mandelbrot}(1967)}]{doi:10.1126/science.156.3775.636}%
  \BibitemOpen
  \bibfield  {author} {\bibinfo {author} {\bibfnamefont {B.}~\bibnamefont
  {Mandelbrot}},\ }\bibfield  {title} {\bibinfo {title} {{How Long Is the Coast
  of Britain? Statistical Self-Similarity and Fractional Dimension}},\ }\href
  {https://doi.org/10.1126/science.156.3775.636} {\bibfield  {journal}
  {\bibinfo  {journal} {Science}\ }\textbf {\bibinfo {volume} {156}},\ \bibinfo
  {pages} {636} (\bibinfo {year} {1967})}\BibitemShut {NoStop}%
\bibitem [{\citenamefont {Bradlyn}\ \emph {et~al.}(2017)\citenamefont
  {Bradlyn}, \citenamefont {Elcoro}, \citenamefont {Cano}, \citenamefont
  {Vergniory}, \citenamefont {Wang}, \citenamefont {Felser}, \citenamefont
  {Aroyo},\ and\ \citenamefont {Bernevig}}]{bradlyn2017topological}%
  \BibitemOpen
  \bibfield  {author} {\bibinfo {author} {\bibfnamefont {B.}~\bibnamefont
  {Bradlyn}}, \bibinfo {author} {\bibfnamefont {L.}~\bibnamefont {Elcoro}},
  \bibinfo {author} {\bibfnamefont {J.}~\bibnamefont {Cano}}, \bibinfo {author}
  {\bibfnamefont {M.}~\bibnamefont {Vergniory}}, \bibinfo {author}
  {\bibfnamefont {Z.}~\bibnamefont {Wang}}, \bibinfo {author} {\bibfnamefont
  {C.}~\bibnamefont {Felser}}, \bibinfo {author} {\bibfnamefont {M.~I.}\
  \bibnamefont {Aroyo}},\ and\ \bibinfo {author} {\bibfnamefont {B.~A.}\
  \bibnamefont {Bernevig}},\ }\bibfield  {title} {\bibinfo {title} {Topological
  quantum chemistry},\ }\href {https://www.nature.com/articles/nature23268}
  {\bibfield  {journal} {\bibinfo  {journal} {Nature}\ }\textbf {\bibinfo
  {volume} {547}},\ \bibinfo {pages} {298} (\bibinfo {year}
  {2017})}\BibitemShut {NoStop}%
\bibitem [{\citenamefont {Po}\ \emph {et~al.}(2017)\citenamefont {Po},
  \citenamefont {Vishwanath},\ and\ \citenamefont {Watanabe}}]{po2017symmetry}%
  \BibitemOpen
  \bibfield  {author} {\bibinfo {author} {\bibfnamefont {H.~C.}\ \bibnamefont
  {Po}}, \bibinfo {author} {\bibfnamefont {A.}~\bibnamefont {Vishwanath}},\
  and\ \bibinfo {author} {\bibfnamefont {H.}~\bibnamefont {Watanabe}},\
  }\bibfield  {title} {\bibinfo {title} {{Symmetry-based indicators of band
  topology in the 230 space groups}},\ }\href
  {https://doi.org/10.1038/s41467-017-00133-2} {\bibfield  {journal} {\bibinfo
  {journal} {Nat. Commun.}\ }\textbf {\bibinfo {volume} {8}},\ \bibinfo {pages}
  {50} (\bibinfo {year} {2017})}\BibitemShut {NoStop}%
\bibitem [{\citenamefont {Cano}\ \emph {et~al.}(2018)\citenamefont {Cano},
  \citenamefont {Bradlyn}, \citenamefont {Wang}, \citenamefont {Elcoro},
  \citenamefont {Vergniory}, \citenamefont {Felser}, \citenamefont {Aroyo},\
  and\ \citenamefont {Bernevig}}]{PhysRevB.97.035139}%
  \BibitemOpen
  \bibfield  {author} {\bibinfo {author} {\bibfnamefont {J.}~\bibnamefont
  {Cano}}, \bibinfo {author} {\bibfnamefont {B.}~\bibnamefont {Bradlyn}},
  \bibinfo {author} {\bibfnamefont {Z.}~\bibnamefont {Wang}}, \bibinfo {author}
  {\bibfnamefont {L.}~\bibnamefont {Elcoro}}, \bibinfo {author} {\bibfnamefont
  {M.~G.}\ \bibnamefont {Vergniory}}, \bibinfo {author} {\bibfnamefont
  {C.}~\bibnamefont {Felser}}, \bibinfo {author} {\bibfnamefont {M.~I.}\
  \bibnamefont {Aroyo}},\ and\ \bibinfo {author} {\bibfnamefont {B.~A.}\
  \bibnamefont {Bernevig}},\ }\bibfield  {title} {\bibinfo {title} {{Building
  blocks of topological quantum chemistry: Elementary band representations}},\
  }\href {https://doi.org/10.1103/PhysRevB.97.035139} {\bibfield  {journal}
  {\bibinfo  {journal} {Phys. Rev. B}\ }\textbf {\bibinfo {volume} {97}},\
  \bibinfo {pages} {035139} (\bibinfo {year} {2018})}\BibitemShut {NoStop}%
\bibitem [{\citenamefont {Benalcazar}\ \emph {et~al.}(2019)\citenamefont
  {Benalcazar}, \citenamefont {Li},\ and\ \citenamefont
  {Hughes}}]{PhysRevB.99.245151}%
  \BibitemOpen
  \bibfield  {author} {\bibinfo {author} {\bibfnamefont {W.~A.}\ \bibnamefont
  {Benalcazar}}, \bibinfo {author} {\bibfnamefont {T.}~\bibnamefont {Li}},\
  and\ \bibinfo {author} {\bibfnamefont {T.~L.}\ \bibnamefont {Hughes}},\
  }\bibfield  {title} {\bibinfo {title} {{Quantization of fractional corner
  charge in ${C}_{n}$-symmetric higher-order topological crystalline
  insulators}},\ }\href {https://doi.org/10.1103/PhysRevB.99.245151} {\bibfield
   {journal} {\bibinfo  {journal} {Phys. Rev. B}\ }\textbf {\bibinfo {volume}
  {99}},\ \bibinfo {pages} {245151} (\bibinfo {year} {2019})}\BibitemShut
  {NoStop}%
\bibitem [{\citenamefont {Po}\ \emph {et~al.}(2018)\citenamefont {Po},
  \citenamefont {Watanabe},\ and\ \citenamefont
  {Vishwanath}}]{PhysRevLett.121.126402}%
  \BibitemOpen
  \bibfield  {author} {\bibinfo {author} {\bibfnamefont {H.~C.}\ \bibnamefont
  {Po}}, \bibinfo {author} {\bibfnamefont {H.}~\bibnamefont {Watanabe}},\ and\
  \bibinfo {author} {\bibfnamefont {A.}~\bibnamefont {Vishwanath}},\ }\bibfield
   {title} {\bibinfo {title} {{Fragile Topology and Wannier Obstructions}},\
  }\href {https://doi.org/10.1103/PhysRevLett.121.126402} {\bibfield  {journal}
  {\bibinfo  {journal} {Phys. Rev. Lett.}\ }\textbf {\bibinfo {volume} {121}},\
  \bibinfo {pages} {126402} (\bibinfo {year} {2018})}\BibitemShut {NoStop}%
\bibitem [{\citenamefont {Schindler}\ \emph {et~al.}(2019)\citenamefont
  {Schindler}, \citenamefont {Brzezi\ifmmode~\acute{n}\else \'{n}\fi{}ska},
  \citenamefont {Benalcazar}, \citenamefont {Iraola}, \citenamefont {Bouhon},
  \citenamefont {Tsirkin}, \citenamefont {Vergniory},\ and\ \citenamefont
  {Neupert}}]{PhysRevResearch.1.033074}%
  \BibitemOpen
  \bibfield  {author} {\bibinfo {author} {\bibfnamefont {F.}~\bibnamefont
  {Schindler}}, \bibinfo {author} {\bibfnamefont {M.}~\bibnamefont
  {Brzezi\ifmmode~\acute{n}\else \'{n}\fi{}ska}}, \bibinfo {author}
  {\bibfnamefont {W.~A.}\ \bibnamefont {Benalcazar}}, \bibinfo {author}
  {\bibfnamefont {M.}~\bibnamefont {Iraola}}, \bibinfo {author} {\bibfnamefont
  {A.}~\bibnamefont {Bouhon}}, \bibinfo {author} {\bibfnamefont {S.~S.}\
  \bibnamefont {Tsirkin}}, \bibinfo {author} {\bibfnamefont {M.~G.}\
  \bibnamefont {Vergniory}},\ and\ \bibinfo {author} {\bibfnamefont
  {T.}~\bibnamefont {Neupert}},\ }\bibfield  {title} {\bibinfo {title}
  {{Fractional corner charges in spin-orbit coupled crystals}},\ }\href
  {https://doi.org/10.1103/PhysRevResearch.1.033074} {\bibfield  {journal}
  {\bibinfo  {journal} {Phys. Rev. Res.}\ }\textbf {\bibinfo {volume} {1}},\
  \bibinfo {pages} {033074} (\bibinfo {year} {2019})}\BibitemShut {NoStop}%
\bibitem [{\citenamefont {Takahashi}\ \emph {et~al.}(2021)\citenamefont
  {Takahashi}, \citenamefont {Zhang},\ and\ \citenamefont
  {Murakami}}]{PhysRevB.103.205123}%
  \BibitemOpen
  \bibfield  {author} {\bibinfo {author} {\bibfnamefont {R.}~\bibnamefont
  {Takahashi}}, \bibinfo {author} {\bibfnamefont {T.}~\bibnamefont {Zhang}},\
  and\ \bibinfo {author} {\bibfnamefont {S.}~\bibnamefont {Murakami}},\
  }\bibfield  {title} {\bibinfo {title} {{General corner charge formula in
  two-dimensional ${C}_{n}$-symmetric higher-order topological insulators}},\
  }\href {https://doi.org/10.1103/PhysRevB.103.205123} {\bibfield  {journal}
  {\bibinfo  {journal} {Phys. Rev. B}\ }\textbf {\bibinfo {volume} {103}},\
  \bibinfo {pages} {205123} (\bibinfo {year} {2021})}\BibitemShut {NoStop}%
\bibitem [{\citenamefont {Kooi}\ \emph {et~al.}(2021)\citenamefont {Kooi},
  \citenamefont {van Miert},\ and\ \citenamefont {Ortix}}]{kooi2021bulk}%
  \BibitemOpen
  \bibfield  {author} {\bibinfo {author} {\bibfnamefont {S.}~\bibnamefont
  {Kooi}}, \bibinfo {author} {\bibfnamefont {G.}~\bibnamefont {van Miert}},\
  and\ \bibinfo {author} {\bibfnamefont {C.}~\bibnamefont {Ortix}},\ }\bibfield
   {title} {\bibinfo {title} {The bulk-corner correspondence of time-reversal
  symmetric insulators},\ }\href
  {https://www.nature.com/articles/s41535-020-00300-7} {\bibfield  {journal}
  {\bibinfo  {journal} {npj Quantum Mater.}\ }\textbf {\bibinfo {volume} {6}},\
  \bibinfo {pages} {1} (\bibinfo {year} {2021})}\BibitemShut {NoStop}%
\bibitem [{\citenamefont {Watanabe}\ and\ \citenamefont
  {Ono}(2020)}]{PhysRevB.102.165120}%
  \BibitemOpen
  \bibfield  {author} {\bibinfo {author} {\bibfnamefont {H.}~\bibnamefont
  {Watanabe}}\ and\ \bibinfo {author} {\bibfnamefont {S.}~\bibnamefont {Ono}},\
  }\bibfield  {title} {\bibinfo {title} {{Corner charge and bulk multipole
  moment in periodic systems}},\ }\href
  {https://doi.org/10.1103/PhysRevB.102.165120} {\bibfield  {journal} {\bibinfo
   {journal} {Phys. Rev. B}\ }\textbf {\bibinfo {volume} {102}},\ \bibinfo
  {pages} {165120} (\bibinfo {year} {2020})}\BibitemShut {NoStop}%
\bibitem [{sup()}]{sup1}%
  \BibitemOpen
  \href@noop {} {}\bibinfo {note} {{See Supplemental Material at URL for
  further details on the filling anomaly for the Hamiltonian in
  Eq.~\eqref{eq:2dsoti}, a proof that the Hamiltonian in Eq.~\eqref{eq:2dsoti}
  is adiabatically connected to the direct sum of two copies of the
  Benalcazar-Bernevig-Hughes models, the formulae to calculate $\Delta n_s$ and
  $\Delta n_t$, the method of the Monte Carlo simulation of the crystal growth,
  the results of the simulation in the case of
  $e^{\mathrm{(n)}}_b=e^{\mathrm{(OAI)}}_b$, and the comparison between the
  obstructed atomic insulator and the normal insulator phases at the same
  $\Delta \mu$.}}\BibitemShut {Stop}%
\bibitem [{\citenamefont {Saito}(1996)}]{saito1996book}%
  \BibitemOpen
  \bibfield  {author} {\bibinfo {author} {\bibfnamefont {Y.}~\bibnamefont
  {Saito}},\ }\href@noop {} {\emph {\bibinfo {title} {{Statistical physics of
  crystal growth}}}}\ (\bibinfo  {publisher} {World Scientific},\ \bibinfo
  {year} {1996})\BibitemShut {NoStop}%
\bibitem [{\citenamefont {Jeong}\ and\ \citenamefont
  {Williams}(1999)}]{jeong1999steps}%
  \BibitemOpen
  \bibfield  {author} {\bibinfo {author} {\bibfnamefont {H.-C.}\ \bibnamefont
  {Jeong}}\ and\ \bibinfo {author} {\bibfnamefont {E.~D.}\ \bibnamefont
  {Williams}},\ }\bibfield  {title} {\bibinfo {title} {{Steps on surfaces:
  experiment and theory}},\ }\href
  {https://www.sciencedirect.com/science/article/abs/pii/S0167572998000107}
  {\bibfield  {journal} {\bibinfo  {journal} {Surf. Sci. Rep.}\ }\textbf
  {\bibinfo {volume} {34}},\ \bibinfo {pages} {171} (\bibinfo {year}
  {1999})}\BibitemShut {NoStop}%
\bibitem [{\citenamefont {Gilmer}\ and\ \citenamefont
  {Bennema}(1972)}]{gilmer_simulation_2003}%
  \BibitemOpen
  \bibfield  {author} {\bibinfo {author} {\bibfnamefont {G.~H.}\ \bibnamefont
  {Gilmer}}\ and\ \bibinfo {author} {\bibfnamefont {P.}~\bibnamefont
  {Bennema}},\ }\bibfield  {title} {\bibinfo {title} {Simulation of {Crystal}
  {Growth} with {Surface} {Diffusion}},\ }\href
  {https://doi.org/10.1063/1.1661325} {\bibfield  {journal} {\bibinfo
  {journal} {J. Appl. Phys.}\ }\textbf {\bibinfo {volume} {43}},\ \bibinfo
  {pages} {1347} (\bibinfo {year} {1972})}\BibitemShut {NoStop}%
\bibitem [{\citenamefont {Huang}(2008)}]{huang2008statistical}%
  \BibitemOpen
  \bibfield  {author} {\bibinfo {author} {\bibfnamefont {K.}~\bibnamefont
  {Huang}},\ }\href@noop {} {\emph {\bibinfo {title} {Statistical mechanics}}}\
  (\bibinfo  {publisher} {John Wiley \& Sons},\ \bibinfo {year}
  {2008})\BibitemShut {NoStop}%
\bibitem [{\citenamefont {Rottman}\ and\ \citenamefont
  {Wortis}(1984)}]{PhysRevB.29.328}%
  \BibitemOpen
  \bibfield  {author} {\bibinfo {author} {\bibfnamefont {C.}~\bibnamefont
  {Rottman}}\ and\ \bibinfo {author} {\bibfnamefont {M.}~\bibnamefont
  {Wortis}},\ }\bibfield  {title} {\bibinfo {title} {{Equilibrium crystal
  shapes for lattice models with nearest-and next-nearest-neighbor
  interactions}},\ }\href {https://doi.org/10.1103/PhysRevB.29.328} {\bibfield
  {journal} {\bibinfo  {journal} {Phys. Rev. B}\ }\textbf {\bibinfo {volume}
  {29}},\ \bibinfo {pages} {328} (\bibinfo {year} {1984})}\BibitemShut
  {NoStop}%
\bibitem [{\citenamefont {Xiao}\ \emph {et~al.}(1988)\citenamefont {Xiao},
  \citenamefont {Alexander},\ and\ \citenamefont
  {Rosenberger}}]{PhysRevA.38.2447}%
  \BibitemOpen
  \bibfield  {author} {\bibinfo {author} {\bibfnamefont {R.-F.}\ \bibnamefont
  {Xiao}}, \bibinfo {author} {\bibfnamefont {J.~I.~D.}\ \bibnamefont
  {Alexander}},\ and\ \bibinfo {author} {\bibfnamefont {F.}~\bibnamefont
  {Rosenberger}},\ }\bibfield  {title} {\bibinfo {title} {{Morphological
  evolution of growing crystals: A Monte Carlo simulation}},\ }\href
  {https://doi.org/10.1103/PhysRevA.38.2447} {\bibfield  {journal} {\bibinfo
  {journal} {Phys. Rev. A}\ }\textbf {\bibinfo {volume} {38}},\ \bibinfo
  {pages} {2447} (\bibinfo {year} {1988})}\BibitemShut {NoStop}%
\bibitem [{\citenamefont {Saito}\ and\ \citenamefont
  {Ueta}(1989)}]{PhysRevA.40.3408}%
  \BibitemOpen
  \bibfield  {author} {\bibinfo {author} {\bibfnamefont {Y.}~\bibnamefont
  {Saito}}\ and\ \bibinfo {author} {\bibfnamefont {T.}~\bibnamefont {Ueta}},\
  }\bibfield  {title} {\bibinfo {title} {{Monte Carlo studies of equilibrium
  and growth shapes of a crystal}},\ }\href
  {https://doi.org/10.1103/PhysRevA.40.3408} {\bibfield  {journal} {\bibinfo
  {journal} {Phys. Rev. A}\ }\textbf {\bibinfo {volume} {40}},\ \bibinfo
  {pages} {3408} (\bibinfo {year} {1989})}\BibitemShut {NoStop}%
\bibitem [{\citenamefont {Mullins}\ and\ \citenamefont
  {Sekerka}(1963)}]{Mullins1963}%
  \BibitemOpen
  \bibfield  {author} {\bibinfo {author} {\bibfnamefont {W.~W.}\ \bibnamefont
  {Mullins}}\ and\ \bibinfo {author} {\bibfnamefont {R.~F.}\ \bibnamefont
  {Sekerka}},\ }\bibfield  {title} {\bibinfo {title} {{Morphological Stability
  of a Particle Growing by Diffusion or Heat Flow}},\ }\href
  {https://doi.org/10.1063/1.1702607} {\bibfield  {journal} {\bibinfo
  {journal} {J. of Appl. Phys.}\ }\textbf {\bibinfo {volume} {34}},\ \bibinfo
  {pages} {323} (\bibinfo {year} {1963})}\BibitemShut {NoStop}%
\bibitem [{\citenamefont {Hilbert}(1891)}]{Hilbert1891459}%
  \BibitemOpen
  \bibfield  {author} {\bibinfo {author} {\bibfnamefont {D.}~\bibnamefont
  {Hilbert}},\ }\bibfield  {title} {\bibinfo {title} {{Ueber die stetige
  Abbildung einer Line auf ein Flächenstück}},\ }\href
  {https://doi.org/10.1007/BF01199431} {\bibfield  {journal} {\bibinfo
  {journal} {Math. Ann.}\ }\textbf {\bibinfo {volume} {38}},\ \bibinfo {pages}
  {459} (\bibinfo {year} {1891})}\BibitemShut {NoStop}%
\bibitem [{\citenamefont {Mandelbrot}\ \emph {et~al.}(1984)\citenamefont
  {Mandelbrot}, \citenamefont {Passoja},\ and\ \citenamefont
  {Paullay}}]{mandelbrot1984fractal}%
  \BibitemOpen
  \bibfield  {author} {\bibinfo {author} {\bibfnamefont {B.~B.}\ \bibnamefont
  {Mandelbrot}}, \bibinfo {author} {\bibfnamefont {D.~E.}\ \bibnamefont
  {Passoja}},\ and\ \bibinfo {author} {\bibfnamefont {A.~J.}\ \bibnamefont
  {Paullay}},\ }\bibfield  {title} {\bibinfo {title} {Fractal character of
  fracture surfaces of metals},\ }\href
  {https://www.nature.com/articles/308721a0} {\bibfield  {journal} {\bibinfo
  {journal} {Nature}\ }\textbf {\bibinfo {volume} {308}},\ \bibinfo {pages}
  {721} (\bibinfo {year} {1984})}\BibitemShut {NoStop}%
\bibitem [{\citenamefont {Imre}(2006)}]{IMRE2006443}%
  \BibitemOpen
  \bibfield  {author} {\bibinfo {author} {\bibfnamefont {A.~R.}\ \bibnamefont
  {Imre}},\ }\bibfield  {title} {\bibinfo {title} {{Artificial fractal
  dimension obtained by using perimeter–area relationship on digitalized
  images}},\ }\href {https://doi.org/https://doi.org/10.1016/j.amc.2005.04.042}
  {\bibfield  {journal} {\bibinfo  {journal} {Appl. Math. Comput,}\ }\textbf
  {\bibinfo {volume} {173}},\ \bibinfo {pages} {443} (\bibinfo {year}
  {2006})}\BibitemShut {NoStop}%
\bibitem [{\citenamefont {Florio}\ \emph {et~al.}(2019)\citenamefont {Florio},
  \citenamefont {Fawell},\ and\ \citenamefont {Small}}]{FLORIO2019551}%
  \BibitemOpen
  \bibfield  {author} {\bibinfo {author} {\bibfnamefont {B.~J.}\ \bibnamefont
  {Florio}}, \bibinfo {author} {\bibfnamefont {P.~D.}\ \bibnamefont {Fawell}},\
  and\ \bibinfo {author} {\bibfnamefont {M.}~\bibnamefont {Small}},\ }\bibfield
   {title} {\bibinfo {title} {{The use of the perimeter-area method to
  calculate the fractal dimension of aggregates}},\ }\href
  {https://doi.org/https://doi.org/10.1016/j.powtec.2018.11.030} {\bibfield
  {journal} {\bibinfo  {journal} {Powder Technol.}\ }\textbf {\bibinfo {volume}
  {343}},\ \bibinfo {pages} {551} (\bibinfo {year} {2019})}\BibitemShut
  {NoStop}%
\bibitem [{\citenamefont {Xu}\ \emph {et~al.}(2024)\citenamefont {Xu},
  \citenamefont {Elcoro}, \citenamefont {Song}, \citenamefont {Vergniory},
  \citenamefont {Felser}, \citenamefont {Parkin}, \citenamefont {Regnault},
  \citenamefont {Ma\~nes},\ and\ \citenamefont {Bernevig}}]{Yuanfeng2024}%
  \BibitemOpen
  \bibfield  {author} {\bibinfo {author} {\bibfnamefont {Y.}~\bibnamefont
  {Xu}}, \bibinfo {author} {\bibfnamefont {L.}~\bibnamefont {Elcoro}}, \bibinfo
  {author} {\bibfnamefont {Z.-D.}\ \bibnamefont {Song}}, \bibinfo {author}
  {\bibfnamefont {M.~G.}\ \bibnamefont {Vergniory}}, \bibinfo {author}
  {\bibfnamefont {C.}~\bibnamefont {Felser}}, \bibinfo {author} {\bibfnamefont
  {S.~S.~P.}\ \bibnamefont {Parkin}}, \bibinfo {author} {\bibfnamefont
  {N.}~\bibnamefont {Regnault}}, \bibinfo {author} {\bibfnamefont {J.~L.}\
  \bibnamefont {Ma\~nes}},\ and\ \bibinfo {author} {\bibfnamefont {B.~A.}\
  \bibnamefont {Bernevig}},\ }\bibfield  {title} {\bibinfo {title}
  {{Filling-enforced obstructed atomic insulators}},\ }\href
  {https://doi.org/10.1103/PhysRevB.109.165139} {\bibfield  {journal} {\bibinfo
   {journal} {Phys. Rev. B}\ }\textbf {\bibinfo {volume} {109}},\ \bibinfo
  {pages} {165139} (\bibinfo {year} {2024})}\BibitemShut {NoStop}%
\bibitem [{\citenamefont {Altland}\ and\ \citenamefont
  {Zirnbauer}(1997)}]{PhysRevB.55.1142}%
  \BibitemOpen
  \bibfield  {author} {\bibinfo {author} {\bibfnamefont {A.}~\bibnamefont
  {Altland}}\ and\ \bibinfo {author} {\bibfnamefont {M.~R.}\ \bibnamefont
  {Zirnbauer}},\ }\bibfield  {title} {\bibinfo {title} {{Nonstandard symmetry
  classes in mesoscopic normal-superconducting hybrid structures}},\ }\href
  {https://doi.org/10.1103/PhysRevB.55.1142} {\bibfield  {journal} {\bibinfo
  {journal} {Phys. Rev. B}\ }\textbf {\bibinfo {volume} {55}},\ \bibinfo
  {pages} {1142} (\bibinfo {year} {1997})}\BibitemShut {NoStop}%
\bibitem [{\citenamefont {Marzari}\ \emph {et~al.}(2012)\citenamefont
  {Marzari}, \citenamefont {Mostofi}, \citenamefont {Yates}, \citenamefont
  {Souza},\ and\ \citenamefont {Vanderbilt}}]{RevModPhys.84.1419}%
  \BibitemOpen
  \bibfield  {author} {\bibinfo {author} {\bibfnamefont {N.}~\bibnamefont
  {Marzari}}, \bibinfo {author} {\bibfnamefont {A.~A.}\ \bibnamefont
  {Mostofi}}, \bibinfo {author} {\bibfnamefont {J.~R.}\ \bibnamefont {Yates}},
  \bibinfo {author} {\bibfnamefont {I.}~\bibnamefont {Souza}},\ and\ \bibinfo
  {author} {\bibfnamefont {D.}~\bibnamefont {Vanderbilt}},\ }\bibfield  {title}
  {\bibinfo {title} {{Maximally localized Wannier functions: Theory and
  applications}},\ }\href {https://doi.org/10.1103/RevModPhys.84.1419}
  {\bibfield  {journal} {\bibinfo  {journal} {Rev. Mod. Phys.}\ }\textbf
  {\bibinfo {volume} {84}},\ \bibinfo {pages} {1419} (\bibinfo {year}
  {2012})}\BibitemShut {NoStop}%
\bibitem [{\citenamefont {Witten}\ and\ \citenamefont
  {Sander}(1981)}]{Witten1981}%
  \BibitemOpen
  \bibfield  {author} {\bibinfo {author} {\bibfnamefont {T.~A.}\ \bibnamefont
  {Witten}}\ and\ \bibinfo {author} {\bibfnamefont {L.~M.}\ \bibnamefont
  {Sander}},\ }\bibfield  {title} {\bibinfo {title} {{Diffusion-Limited
  Aggregation, a Kinetic Critical Phenomenon}},\ }\href
  {https://doi.org/10.1103/PhysRevLett.47.1400} {\bibfield  {journal} {\bibinfo
   {journal} {Phys. Rev. Lett.}\ }\textbf {\bibinfo {volume} {47}},\ \bibinfo
  {pages} {1400} (\bibinfo {year} {1981})}\BibitemShut {NoStop}%
\bibitem [{\citenamefont {Witten}\ and\ \citenamefont
  {Sander}(1983)}]{Witten1983}%
  \BibitemOpen
  \bibfield  {author} {\bibinfo {author} {\bibfnamefont {T.~A.}\ \bibnamefont
  {Witten}}\ and\ \bibinfo {author} {\bibfnamefont {L.~M.}\ \bibnamefont
  {Sander}},\ }\bibfield  {title} {\bibinfo {title} {{Diffusion-limited
  aggregation}},\ }\href {https://doi.org/10.1103/PhysRevB.27.5686} {\bibfield
  {journal} {\bibinfo  {journal} {Phys. Rev. B}\ }\textbf {\bibinfo {volume}
  {27}},\ \bibinfo {pages} {5686} (\bibinfo {year} {1983})}\BibitemShut
  {NoStop}%
\end{thebibliography}
\end{document}